%% file: tex16743.tex
\documentclass[structabstract]{aa}
\usepackage{subfig}
\usepackage{txfonts} 
\usepackage{color}
\usepackage{graphicx}
\usepackage{natbib} 
\bibpunct{(}{)}{;}{a}{}{,} 
\usepackage {amssymb}
\usepackage[english]{babel}
\usepackage{natbib}
\usepackage{rotating}

\title{Sequential star formation in IRAS 06084-0611 (GGD 12-15) 
\subtitle{From intermediate-mass to high-mass stars\thanks{ Based on observations collected at the European Southern Observatory at Paranal, Chile (ESO program 078.C-0780)}}}

\author{K. M. Maaskant\inst{1,}\inst{2} \and A. Bik\inst{3} \and L.B.F.M. Waters\inst{4,}\inst{1}  \and L. Kaper \inst{1} \and Th. Henning\inst{3} \and E. Puga\inst{5,}\inst{6} \and M. Horrobin\inst{7} \and J. Kainulainen\inst{3} }
\institute{Astronomical Institute Anton Pannekoek, University of Amsterdam, P.O. Box 94249, 1090 GE Amsterdam, The Netherlands \and Leiden Observatory, Leiden University, P.O. Box 9513, 2300 RA Leiden, The Netherlands \and Max-Planck-Institut f\"ur Astronomie, K\"onigstuhl 17, 69117 Heidelberg, Germany \and SRON-Netherlands Institute for Space Research, Sorbonnelaan 2, 3584 CA Utrecht, the Netherlands \and Centro de Astrobiolog\'ia (CSIC-INTA), 28850 Torrej\'on de Ardoz, Madrid, Spain \and Instituut voor Sterrenkunde, Celestijnenlaan 200D, B-3001 Leuven, Belgium \and I.Physikalisches Institut, Universit\"at zu K\"oln, 50937 K\"oln, Germany}

\date{\today}

\abstract{The formation and early evolution of high- and intermediate-mass stars towards the main sequence involves the interplay of stars in a clustered and highly complex environment. To obtain a full census of this interaction, the Formation and Early evolution of Massive Stars (FEMS) collaboration studies a well-selected sample of 10 high-mass star-forming regions.} 
{In this study we examine the stellar content of the high-mass star-forming region centered on IRAS 06084-0611 in the Monoceros R2 cloud.}
{Using the near-infrared \textit{H-} and \textit{K}-band spectra from the VLT/SINFONI instrument on the ESO \emph{Very Large Telescope} (VLT)and photometric near-infrared NTT/SOFI, 2MASS and \emph{Spitzer}/IRAC data, we were able to determine the spectral types for the most luminous stars in the cluster.}
{Two very young and reddened massive stars have been detected by SINFONI: a massive Young Stellar Object (YSO) consistent with an  early-B spectral type and a Herbig Be star. Furthermore, stars of  spectral type G and K are detected while still in the Pre-Main Sequence (PMS) phase. We derive additional properties such as temperatures, extinctions, radii and masses. We present a Hertzsprung-Russell diagram and find most objects having intermediate masses between $\sim$1.5-2.5 M$_{\odot}$. For these stars we derive a median cluster age of $\sim$4 Myr. } 
{Using \emph{Spitzer}/IRAC data we confirm earlier studies that the younger class 0/I objects are centrally located  while the class II objects are spread out over a larger area, with rough scale size radii of $\sim$0.5 pc and  $\sim$1.25 pc respectively. Moreover, the presence of a massive YSO, an ultracompact H~{\sc ii} region and highly reddened objects in the center of the cluster suggest a much younger age of $<$ 1 Myr. A possible scenario for this observation would be sequential star formation along the line of sight; from a cluster of intermediate-mass to high-mass stars.
 }


\begin{document}

\maketitle

\section{Introduction}

\label{intro}

\begin{figure*}
\centering
\includegraphics[scale= 0.53 ]{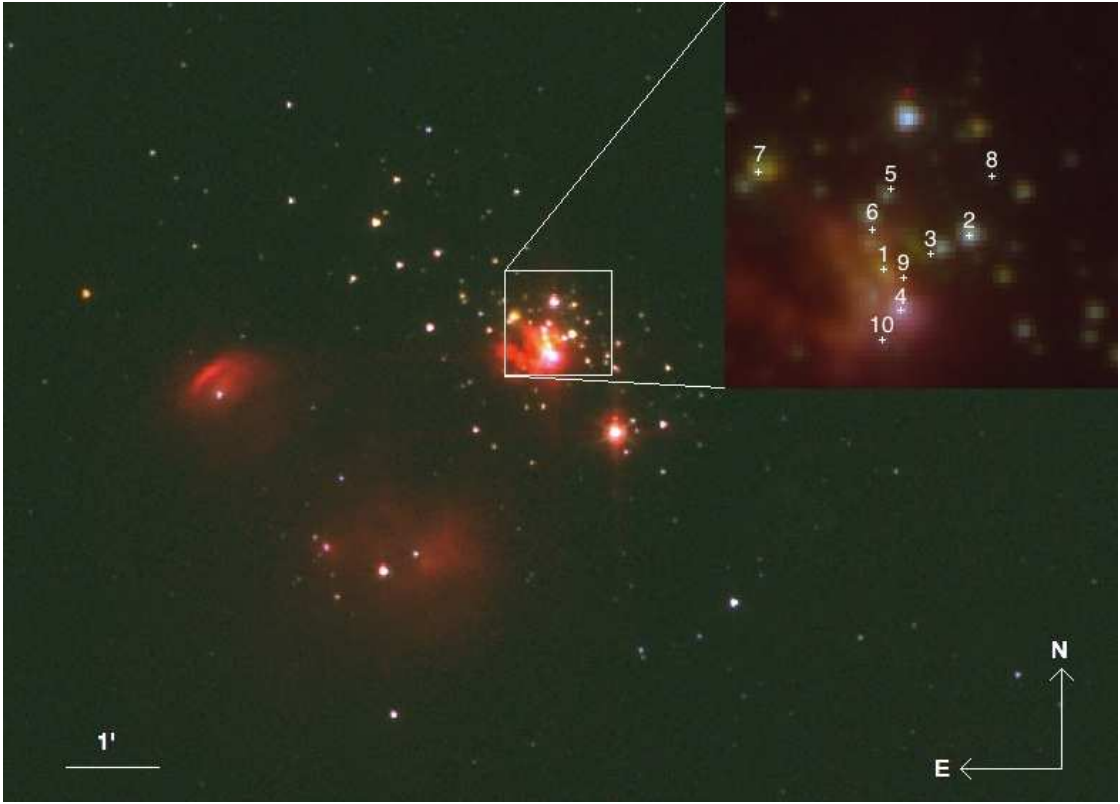} 
\caption{\label{fig:morphology_IR_spitzer_color} \emph{Spitzer}/IRAC three-color image of IRAS 06084-0611 and its surroundings (blue: 3.6 $\mu$m, green: 4.5 $\mu$m, red: 5.8 $\mu$m). The dense cluster region delimited by the white box is the area covered by the SINFONI observations. The right upper inlay figure is a close-up of the SINFONI region. The positions of the VLA radio sources and their respective ID numbers \citep{2002Gomezetal} are marked with plus signs. Note that VLA 1 represents an H~{\sc ii} region and is also visible in the IRAC bands. The field of view is equal to Figure \ref{fig:morphology_dss_IR_spitzer}. }
\end{figure*}

The formation of stars and planets is a result of complex intertwined behavior of gas and dust, strongly depending on the initial conditions. The presence of high-mass stars is believed to play a major role in the process of star formation in clusters \citep{2007ZinneckerYorke}. Since high-mass stars preferentially form in clusters, they inevitably exert influence on the cluster surroundings. Massive stars can clear out their direct surroundings through their strong winds and ionizing fluxes more rapidly than less massive stars. The energy they release in the interstellar medium powers H~{\sc ii}  regions and bubbles and is thought to trigger second-generation star formation events. 

Several mechanisms have been proposed to explain triggered star formation. The collect and collapse model (\citealt{1977ElmegreenLada}) involves the propagation of ionization and shock fronts through a molecular cloud complex. Between these fronts a dense shell of material is trapped and when grown large enough, global shell fragmentation occurs and new stars are formed. Alternatively, triggered star formation can occur in pre-existing clumps. Through dynamical processes such as the expansion of an H~{\sc ii}  region in a non-homogeneous neutral medium, high pressure accumulates the dense material into layers, shells or rings which then gravitationally collapse \citep{1998Elmegreen}.  We refer to \citet{2010Lada} and \citet{2009Andreetal} and references therein for reviews of triggered star formation scenarios. 

Infrared (IR), sub-millimeter and radio imaging has allowed these earliest stages of protostellar evolution to become accessible to observers. Observational studies have started to map the interaction of high-mass stars and previously hidden stellar content in those star-forming regions (\citealt{2000Blumetal}; \citealt{2004Bik}; \citealt{2004Alvarezetal}; \citealt{2006Churchwelletal}). Much less ambiguous identification and classification of the stellar content has been demonstrated by the diagnostic power of near-infrared (NIR) spectroscopy \citep{2002Hansonetal,2005Biketal,2010Pugaetal,2010Martinsetal,2010Biketal}.

This paper is the third in a program to obtain a full NIR spectroscopic census of the stellar content of a sample of high-mass star-forming regions by means of integral field spectroscopy using VLT/SINFONI. 
The first paper by \citet{2010Biketal} studies the star formation process in the high-mass star-forming region RCW 34. A spatial distinction between class II sources inside a `bubble region'  on one hand, and the H~{\sc ii}  region, the massive stars and the class 0/I sources in the molecular cloud on the northern edge of the bubble on the other hand, suggest that sequential star formation is at work from south to north. The second paper by \citet{2011Wangetal} examines star formation in the S255 complex: three regions of massive star formation at different evolutionary stages that reside in the same natal molecular cloud. It was found that the age difference between the older low-mass cluster and the young massive cores indicates different stellar populations in the cluster. \citet{2011Wangetal} propose the triggered outside-in collapse star formation scenario for this region.

These earlier papers show that the interaction between the young massive stars and the surrounding molecular cloud could not only lead to the destruction of the cloud, but also to the formation of a younger generation of stars. Multiple generations of newly formed stars are identified and star formation scenarios have been proposed in which clusters of stars could have triggered the collapse of a pre-existing dense core in the molecular clouds. It seems that the interaction between stars and the surrounding molecular cloud could lead to the formation of a younger generation including massive stars. In a similar way, this paper studies the interplay between the cluster environment of IRAS 06084-0611 and the embedded young stars within it.

The star-forming region IRAS 06084-0611, often referred to as GGD 12-15, is deeply embedded in the Monoceros molecular cloud at a distance of  830 $\pm$ 50 pc \citep{1976HerbstRacine}. Figure \ref{fig:morphology_IR_spitzer_color} gives an overview of the cluster surroundings in the first three IRAC bands. We can distinguish the \emph{Spitzer} image in roughly three regions: eastward we observe the signature of a bow shock; gas flowing out of the central star colliding with the ISM. In the south we observe an extended emission region showing (very) weak filamentary structures. But the main focus of this paper is the dense cluster region covered by SINFONI, delimited by the white box. In that region, active signs of star-formation have been observed from infrared to radio wavelengths: an H~{\sc ii}  region, a Herbig Be star, a massive YSO, extended CO-bipolar outflows, an H$_2$O maser and about a dozen potential T Tauri objects. 

\emph{Radio observations:} Radio continuum observations at 3.6 and 6 cm were obtained by \cite{2000Gomezetal, 2002Gomezetal} using the Very Large Array (VLA). Around the bright cometary H~{\sc ii}  region VLA 1, nine extremely compact ($\leq$0.3'') and faint radio sources have been found (Figure \ref{fig:morphology_IR_spitzer_color}). \citet{2002Gomezetal} estimate that VLA 1 is excited by a B0.5 ZAMS star with a luminosity of $\sim$1.0 $\times$ 10$^4$ $L_{\odot}$. The derived spectral index for VLA 7 (+0.6 $\pm$ 0.3) fits the stellar wind model better than other possibilities, supporting the idea that VLA 7 is the powering source of the molecular CO outflow which extends $\sim$6$\arcmin$ from SE to NW \citep{2008Qinetal}. The remaining radiosources show variability at radio wavelengths, and their spectral indices are characteristically negative. It has been suggested that their emission mechanism could be explained by gyro-synchrotron radiation from the active magnetospheres of T Tauri stars. 

\emph{IR observations:} Several infrared studies covering a range from 1 to 120 $\mu$m (\citealt{2004Bik}; \citealt{2005Biketal}; \citealt{2006Biketal}; \citealt{1983ReipurthWamsteker}; \citealt{1985OlofssonKoornneef}; \citealt{1985Harveyetal}; \citealt{1994Hodapp}) have revealed about a dozen objects. It has been suggested that most of these objects are PMS cluster members. Mid-infrared observations at  8.7, 9.7 and 12.5 $\mu$m \citep{2003PersiTapia} revealed two objects: one identified with the compact cometary H~{\sc ii}  region VLA 1 \citep{2002Gomezetal} and the second with the faint and unresolved radio continuum source VLA 4. The latter has been identified as a Herbig Be star since the IR counterpart shows large IR excess and its derived IR luminosity is $L_{1 - 20 \mu m}$ = 350 $L_{\odot}$, indicating the presence of a young massive star. The spectrum of VLA 4  between 2.47 and 11.62 $\mu$m shows the presence of IR emission bands at 3.3, 6.2, 7.7, 8.6 and 11.2 $\mu$m also known as Polycyclic Aromatic Hydrocarbon (PAH) features (\citealt{2003PersiTapia}; \citealt{2009Boersmaetal}).

\emph{Molecular line observations:}  The most recent study of the CO outflow has been carried out by \citet{2008Qinetal} in which CO J = 3-2 and J = 2-1 lines are mapped. VLA7 is located centrally on the axis of the outflow and is a likely candidate to be the powering source. Moreover, the origin and position of the central source powering the outflow is studied by \citet{2008Satoetal} using NIR imaging polarimetry and they confirm that the central object indeed seems to coincide with VLA 7. High density HCO$^+$ gas \citep{1988Heatonetal}, a H$_2$O maser \citep{1995Tofanietal} and a NH$_3$ region \citep{1989Torrellesetal} have been observed in the center of the outflow. There are two maxima in the ammonia region suggesting that the structure could represent a molecular toroid which helps to collimate the outflow. 

The objective of this paper is to probe the star formation history of the cluster. In the next section, the spectroscopic VLT/SINFONI and supporting photometric NTT/SOFI, 2MASS and \emph{Spitzer}/IRAC observations are presented. The analysis is divided in three parts. First, the photometric colors are presented and an overview is given of the cluster surrounding. Second, using the \textit{H-} and \textit{K}-band spectra, we present spectral classifications of the high- and intermediate-mass stellar population and determine their properties (e.g. extinction, age, mass).  And third, the objects are placed in a Hertzsprung Russell diagram and the cluster age is estimated. Finally, we provide an interpretation of the formation history of IRAS 06084-0611. The last section summarizes the main conclusions of the paper.

\section{Observations and data reduction}

 \begin{figure}

\subfloat[]{
\includegraphics[width= 0.48\textwidth ]{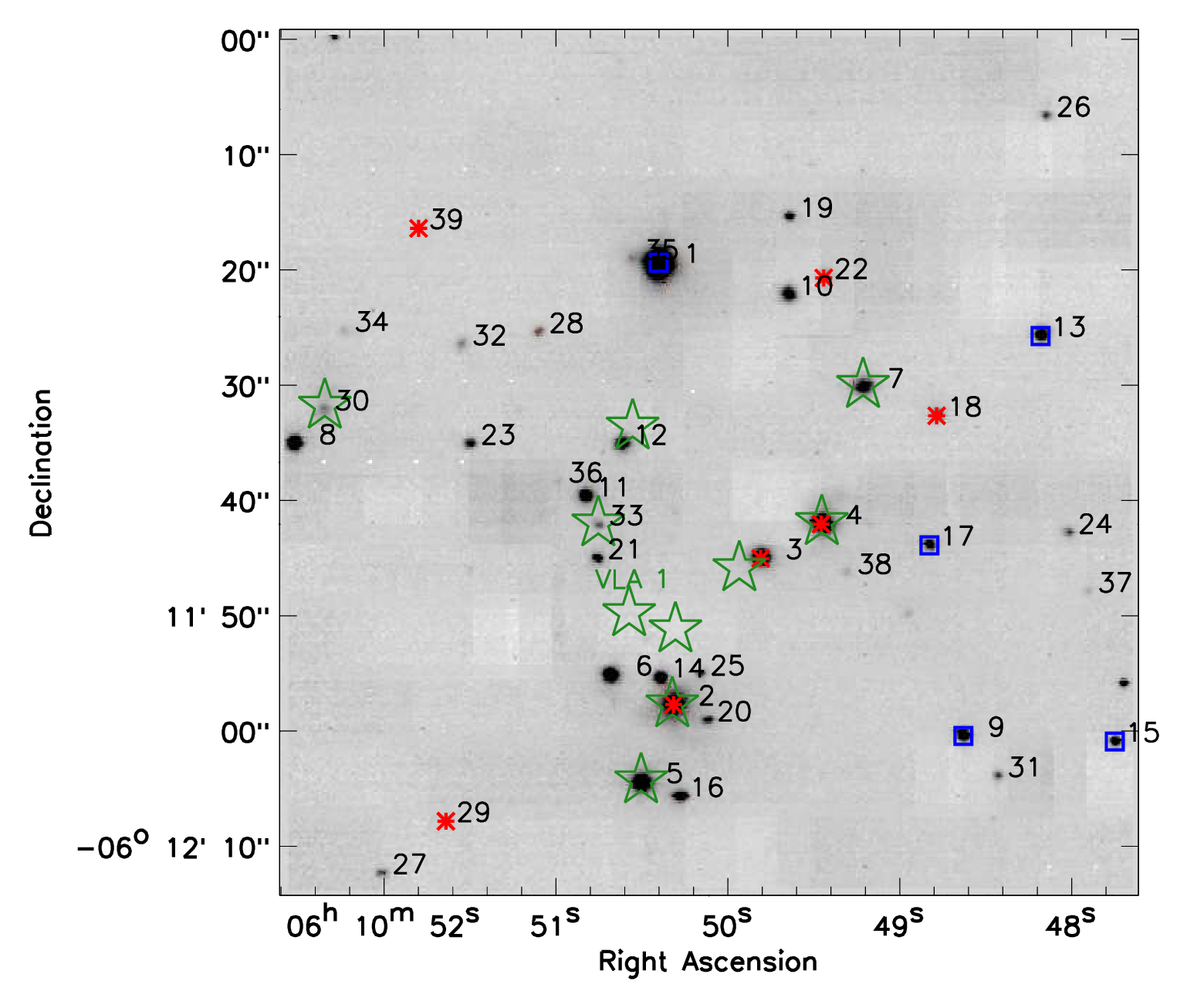} 
\label{fig:objects_VLA}
}
\\ \subfloat[]{
\includegraphics[width=0.48\textwidth]{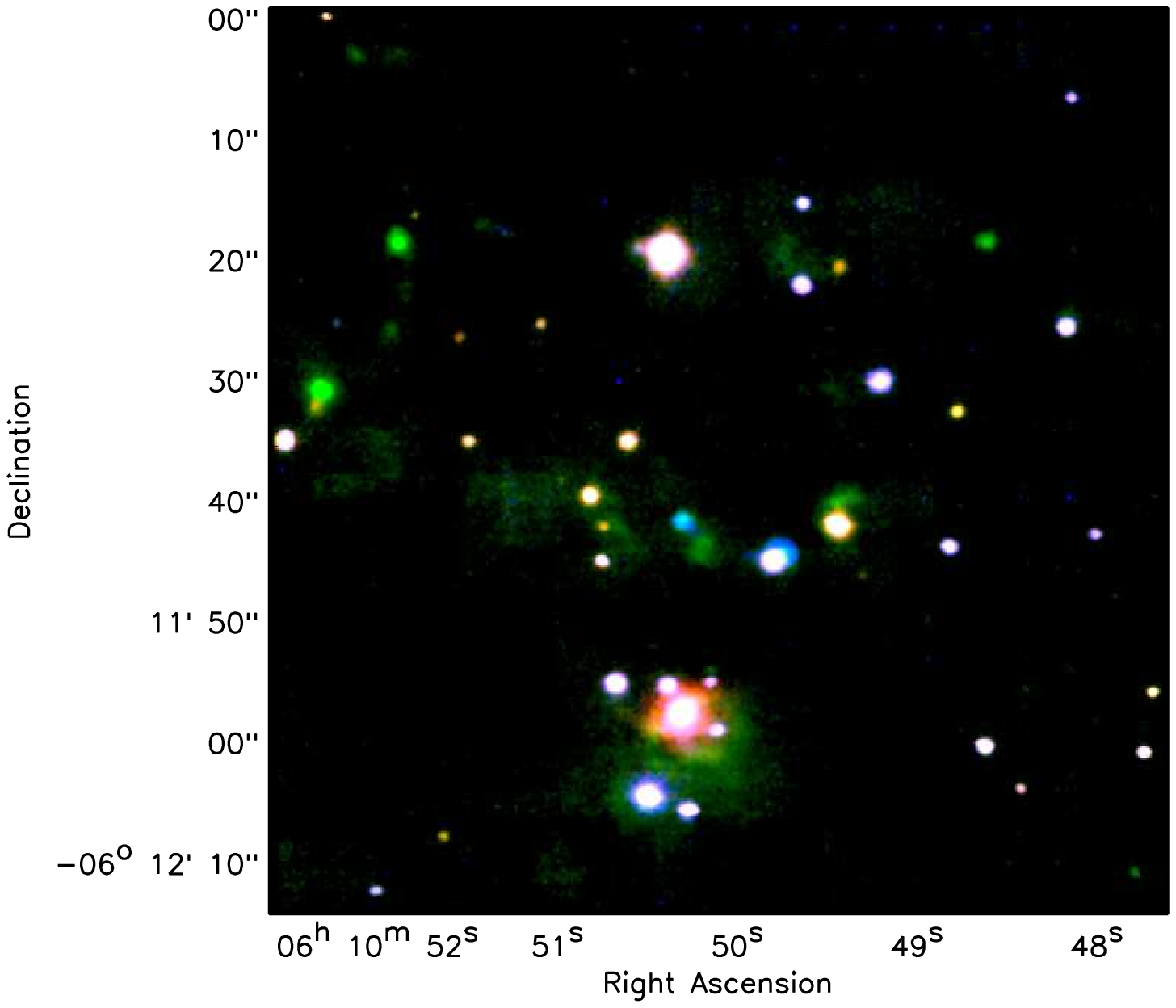}
\label{fig:subSINFONI} 
}
\caption{\label{fig:SINFONI_field}VLT/SINFONI observation of IRAS 06084-0611.  \textit{Top:} Sinfoni 2.16 $\mu$m line-map with the object identification numbers sorted by SOFI \textit{K}-band brightness. The positions of evolutionary class \citep{1987Lada} 0/I (red asterisks), class II (blue squares) objects and VLA radio detections (green stars, \citealt{2002Gomezetal}, see Figure \ref{fig:morphology_IR_spitzer_color} for all their respective ID numbers) are marked. We do not detect the NIR counterpart of VLA 1 in the SINFONI data. \textit{Bottom:} Three-color image created from the SINFONI line maps with green: H$_2$ (2.12 $\mu$m, shocked gas), blue: [Fe~{\sc ii}] (1.644 $\mu$m, shocked outflows) and red: Br-$\gamma$ (2.16 $\mu$m, ionized gas). Green  H$_2$ and blue [Fe~{\sc ii}] spots are clearly visible, suggesting the presence of outflows and jets, respectively, and thereby tracing active star formation.  (Epoch: J2000)}
\end{figure}

\begin{figure}[]
\centering
\subfloat[]{
\includegraphics[scale= 0.5, angle = 180]{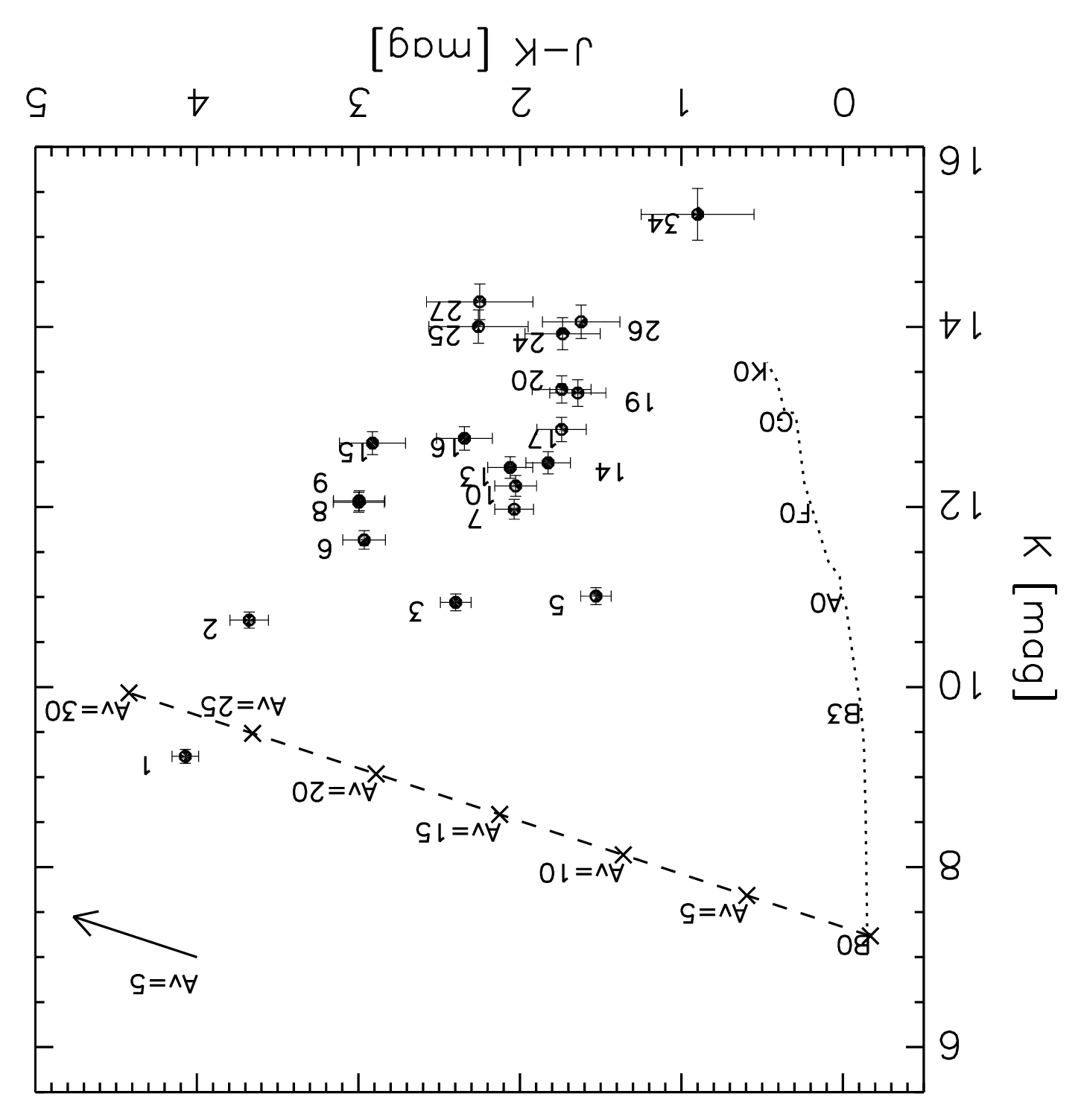}
\label{fig:color_SOFI_jkk_match}
}
\\ \subfloat[]{
\includegraphics[scale= 0.5, angle = 180]{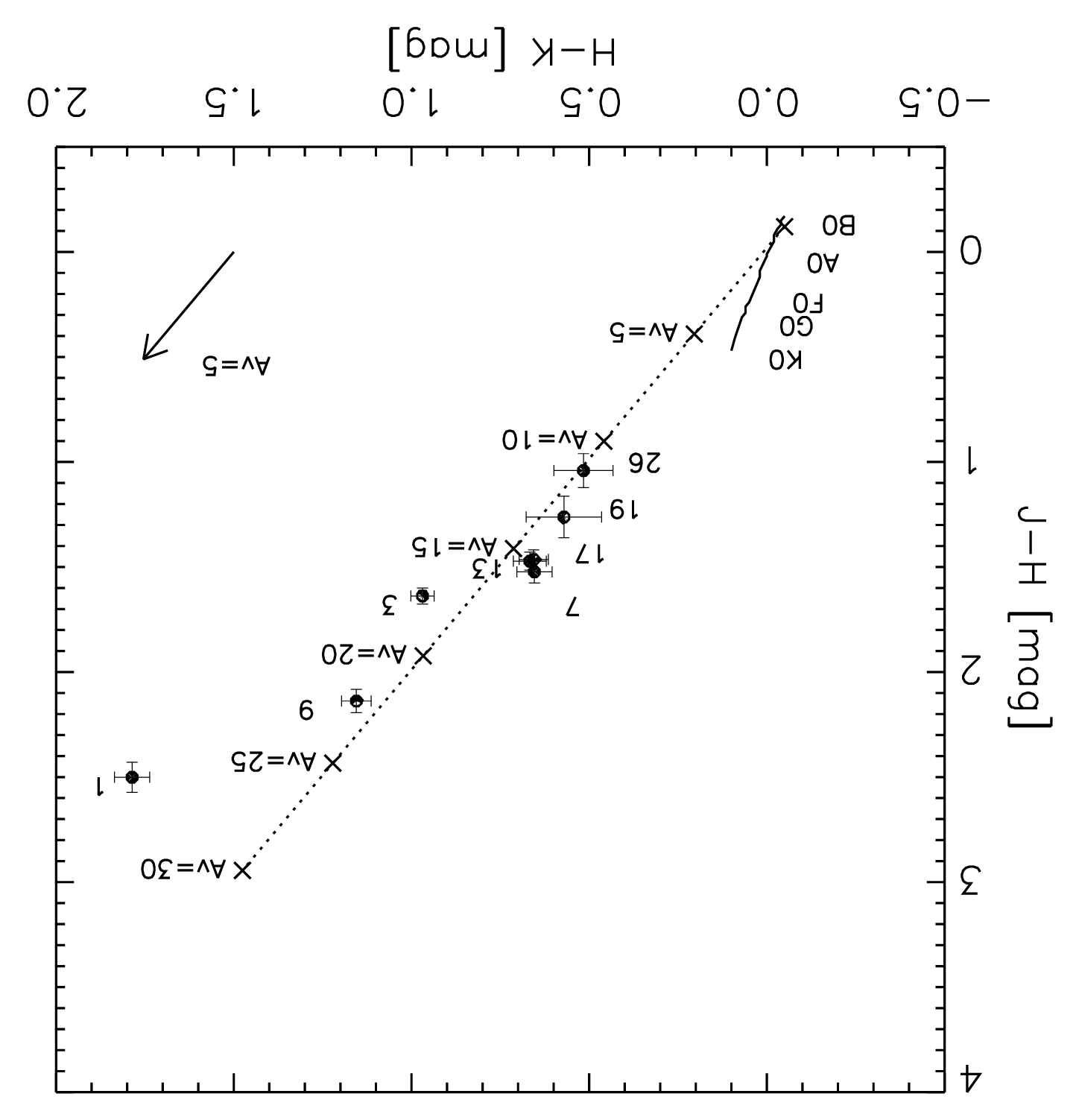}
\label{fig:color_2MASS_hkjh_match}
}
\caption{\label{fig:IR_colors}\textit{Top:} the SOFI (\textit{J}-\textit{K}, \textit{K}) diagram. \textit{Bottom:}  the 2MASS (\textit{H-}\textit{K}, \textit{J}-\textit{H}) diagram. Both figures only show the detections in the SINFONI field. The visual extinction A$_{\rm V}$ is calculated using the extinction law of \cite{2006ChiarTielens}. The isochrones of main sequence dwarf stars (\citealt{2000Blumetal}; \citealt{1988BesselBrett}) are plotted with an adopted distance for  IRAS 06084-0611 of 830 $\pm$ 50 pc \citep{1976HerbstRacine}. Note that narrow-band filters were used with SOFI observations. }
\end{figure}

\begin{figure}[]
\centering

\subfloat[]{
\includegraphics[scale= 0.5]{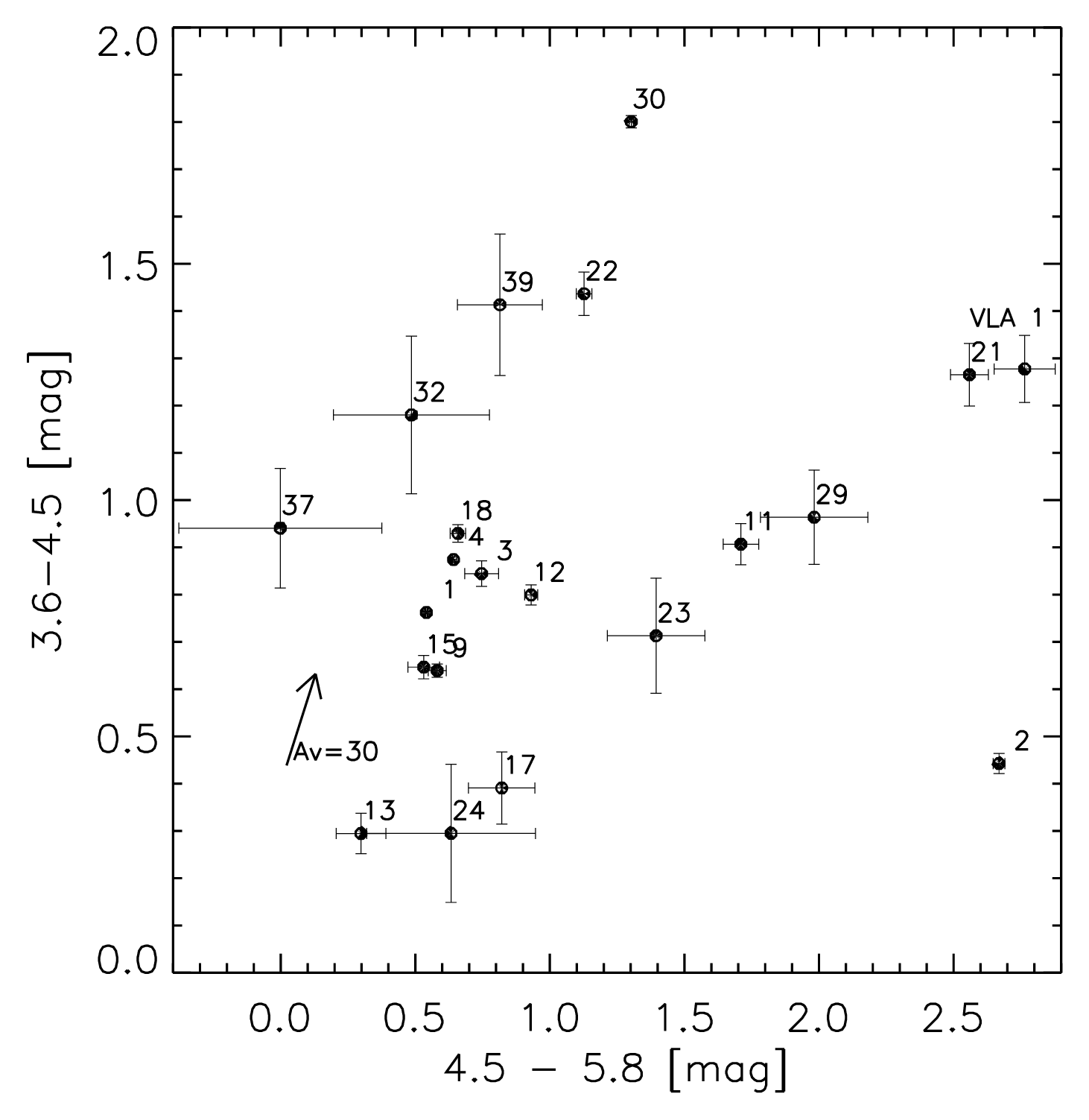}
\label{fig:color_SPITZER_3bands_match}
}
\\ \subfloat[]{
\includegraphics[ scale=0.5, angle = 180]{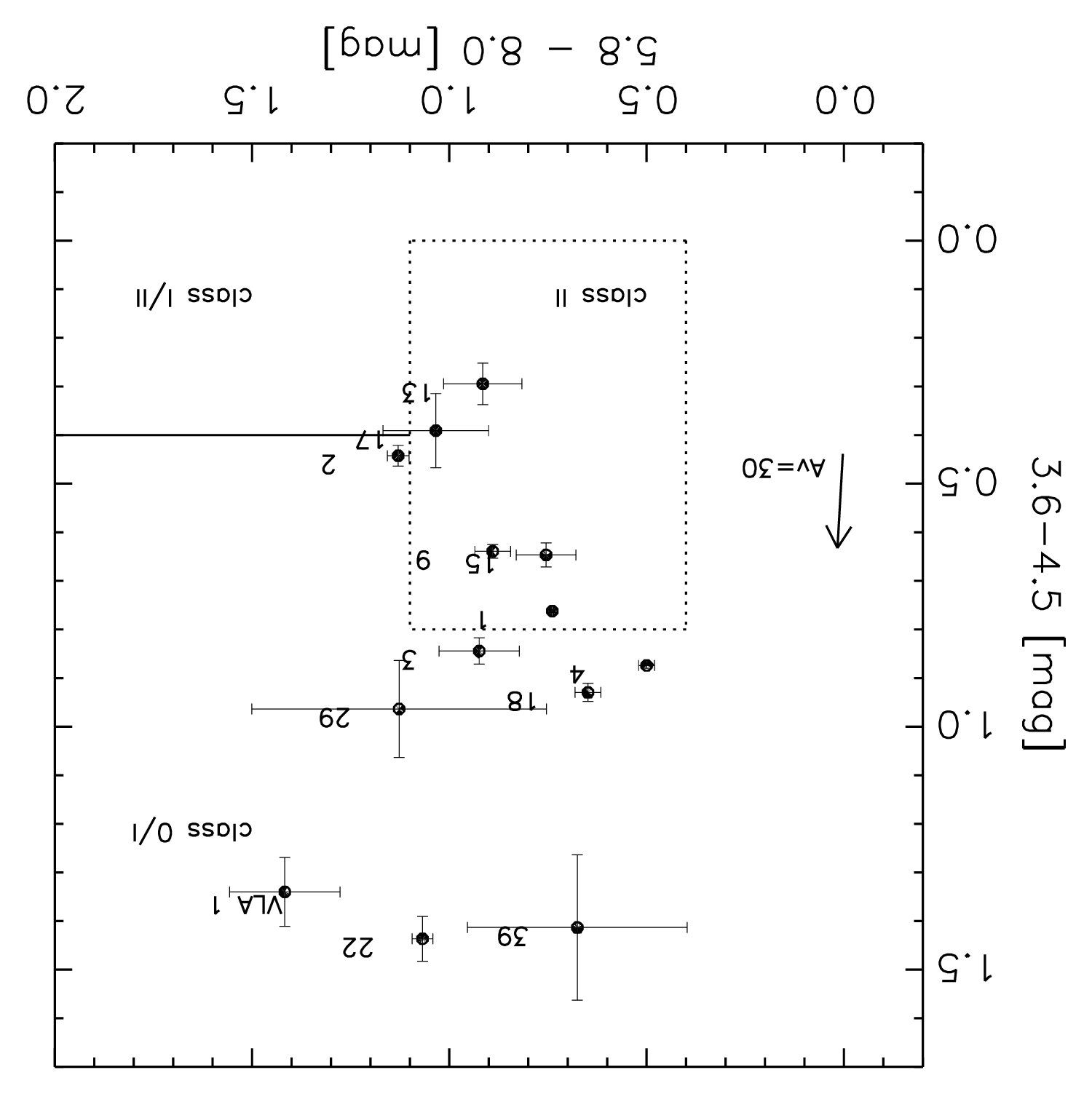}
\label{fig:color_spitzer}
}

\caption{\label{fig:color_spitzer_both} \textit{Top:}  \emph{Spitzer}/IRAC (3.6 - 4.5,  4.5 - 5.8) $\mu$m diagram of objects in the SINFONI field centered on IRAS 06084-0611. Photospheric colors from main-sequence stars are usually located around (0, 0). The extinction vector has been calculated using the extinction law of \cite{2006ChiarTielens}.  \textit{Bottom:} \emph{Spitzer} (3.6 - 4.5,  5.8 - 8.0) $\mu$m diagram. We used \citealt{2004Allenetal} and \citealt{2004Megeathetal} to specify the location of  class 0/I, class I/II and class II sources (dotted lines). }
\end{figure}

\subsection{Spectroscopy: VLT/SINFONI \textit{H-} and \textit{K}-band observations}

Spectroscopic near-infrared \textit{H-} and \textit{K}-band observations were taken using the Integral Field spectrograph SINFONI (\citealt{2003Eisenhaueretal}; \citealt{2004Bonnetetal}) on the Unit Telescope 4 (YEPUN) of the ESO Very Large Telescope at Paranal, Chile (see Figure \ref{fig:objects_VLA} for the continuum image of the region). The observations were performed in service mode on January 30 and February 4, 2007.   The observations were performed using the non-AO mode providing a 8$\arcsec$ $\times$ 8$\arcsec$ field of view.  The spectrograph was operated using the \textit{H}+\textit{K} grating providing a spectral resolving power of R=1500 in the \textit{H-} and \textit{K}-band in a single exposure. The observations are centered on RA= \mbox{06$^{h}$10$^{m}$50.14$^{s}$}, Dec = \mbox{-06$^{\circ}$11$\arcmin$ 36.82$\arcsec$} (J2000) and cover a total area of 74$\arcsec$ $\times$ 74$\arcsec$.  To cover the entire cluster we applied a mapping pattern with 4$\arcsec$ offsets in RA and 6.75$\arcsec$ offsets in Dec to cover every position in the cluster at least twice with a detector integration time (DIT) of 30 seconds to guarantee a good S/N ($\sim$ 70) for the early type stars in the cluster. The average seeing during the observations was $\sim$ 0.8$\arcsec$. 
Furthermore, every 3 minutes, a sky frame was taken using the same DIT as the science observations. The sky positions were chosen based on existing NIR imaging in order to avoid contamination. Immediately after every science observation, a telluric standard of B spectral type was observed, matching as closely as possible the airmass of the object. This star was used for the removal of the telluric absorption lines as well as for flux calibration. 

The data reduction was performed using the SPRED software package version 1.37 (\citealt{2004Schreiber}; \citealt{2006Abuter}). The data reduction consists of correcting for dark, flat-field and distortion as well as reconstructing the final 3D data cubes \citep[see][for a detailed description]{2010Biketal}.  The final reduced SINFONI data of IRAS 06084-0611 consist of 3D data cubes with two spatial (x-y) dimensions and a third dimension encompassing  the \textit{H-} and \textit{K}-band (1.4 - 2.45 $\mu$) spectra for each individual pixel. The VLT/SINFONI line $+$ continuum image is shown on Figure \ref{fig:subSINFONI}.

\begin{table*}[!ht]
\begin{center}
\caption{\label{tab:magnitudes}\normalsize{\textsc{Photometric properties of objects located in the SINFONI field of view.}}}

\scriptsize{
\hspace{0cm}\begin{tabular}{c c c c c c c c c c}
\hline
\hline
\multicolumn{1}{c}{ }& \multicolumn{2}{c}{\textit{NTT/SOFI}}& \multicolumn{3}{c}{\textit{2MASS}}& \multicolumn{4}{c}{\textit{Spitzer/IRAC}} \\
Object& \textit{J} & \textit{K}   & \textit{J}  & \textit{H}  & \textit{K}   & band 1 & band 2 & band 3 & band 4 \\
\#  & [1.25 $\mu$m] & [2.16 $\mu$m]  &[1.25 $\mu$m]  &[1.65 $\mu$m]   & [2.16 $\mu$m]  & [3.6 $\mu$m] & [4.5 $\mu$m] & [5.8 $\mu$m] & [8.0 $\mu$m] \\
\hline
\input{tab1_16743.tex}
\hline

\end{tabular}
}

\end{center}
\end{table*}%

\begin{figure*}
\centering
\includegraphics[scale= 0.7 ]{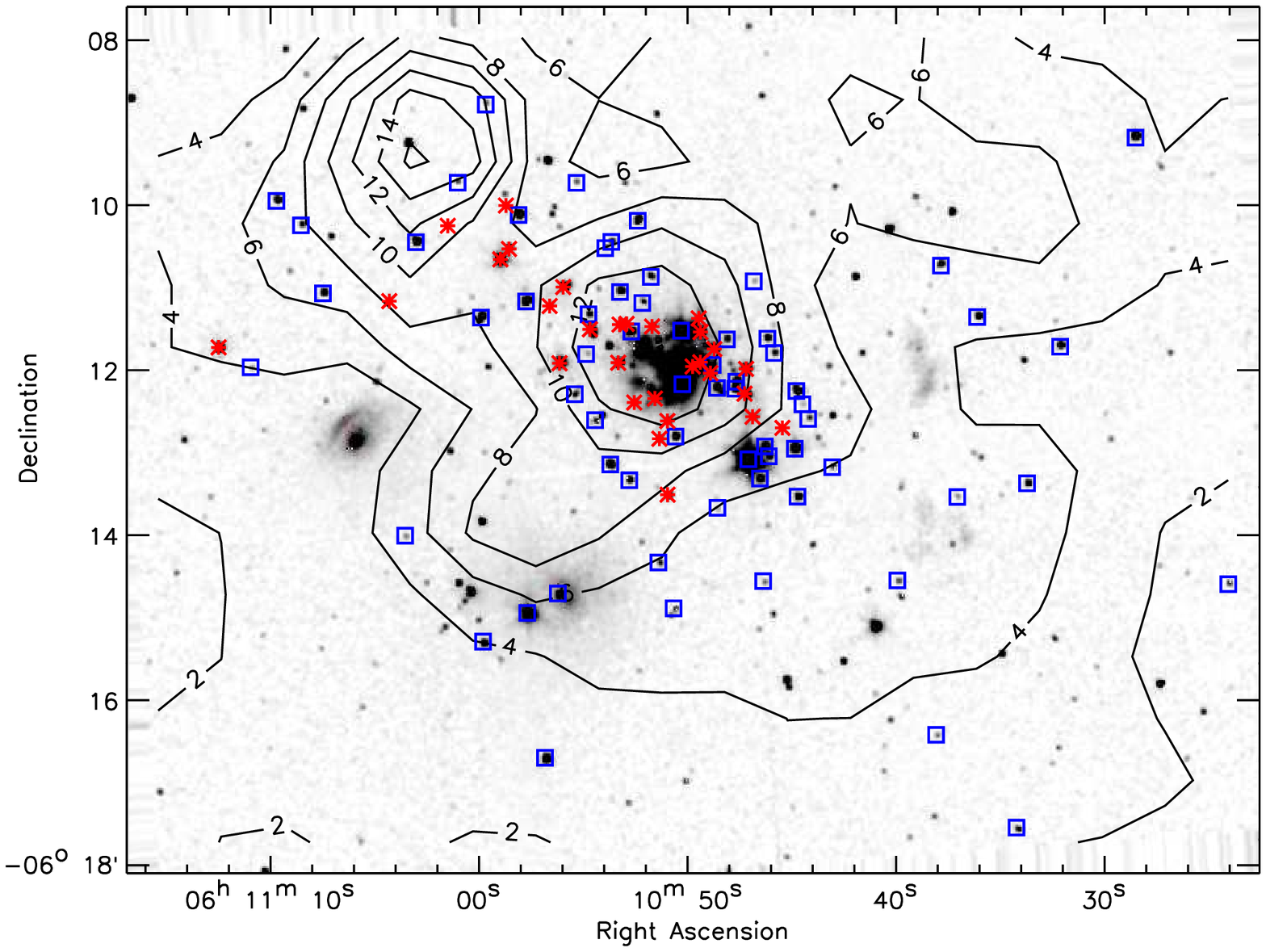} 
\caption{\label{fig:morphology_dss_IR_spitzer}  Plotted on top of the IRAC 4.5 $\mu$m band image of IRAS 06084-0611: the positions of evolutionary class \citep{1987Lada} 0/I (red asterisks) and class II (blue squares) objects and an extinction map based on combination of 2MASS and SOFI photometry. The contours are in levels of A$_{\rm V}$. Note that the class II objects are spread out over a larger region compared to the younger class 0/I objects. (Epoch: J2000). }
\end{figure*}

\subsection{Photometry: \emph{Spitzer}/IRAC, 2MASS and NTT/SOFI observations}

\label{sec:NTTSOFI}

Complementary basic calibration archival \emph{Spitzer}/IRAC data \citep{2004Fazioetal} are used to study the region as well as its surroundings at longer wavelengths. The IRAC data covering IRAS 06084-0611 (PID: 6, PI: Fazio) were processed using the standard pipeline version S14.0.0. The data were downloaded from the archive and processed using the MOPEX software version 16.2.1 \citep{2006Makovoz}. \emph{Spitzer}/IRAC photometry of IRAS 06084-0611 has previously been studied in \citet{2009Gutermuth}.

Using IRAF, photometry was applied with an aperture of 1.5 $\times$ FWHM (with a FWHM $\sim$ 1.7 pixels for all bands), a sky annulus of 3.5 $\times$ FWHM and an annulus width of 2.5 $\times$ FWHM. To obtain the magnitudes of the objects, the appropriate aperture corrections and zeropoints were adopted from the \emph{Spitzer}/IRAC website.

Additionally, 2MASS \citep{2006Skrutskieetal} data was used in which detections of poor data quality were discarded as well as SOFI data taken from \citet{2004Bik}, which covers a 5$\arcmin$ $\times$ 5$\arcmin$ field around IRAS 06084-0611 using narrow-band continuum filters in \textit{J} and \textit{K}.  An advantage of narrow-band filters is that the observations are better representatives of the stellar continuum flux in the \textit{J-} and \textit{K}-band as they were chosen not to include emission lines.  The \textit{J-}band continuum is centered on 1.215 $\mu$, with a width of 0.018 $\mu$m and the \textit{K-}band is centered on 2.09 $\mu$m with a width of 0.02 $\mu$m. For comparison with 2MASS broadband photometry and isochrone colors of standard main sequence stars, a correction was applied to address the observed differences between narrow-band and broad-band filters \citep{2004BikThesis} Finally, when a SINFONI source coincides in position within 1.25\arcsec with a 2MASS, SOFI or IRAC source the detections were matched. The matching procedure connected 39 objects with SOFI and SINFONI detections, and available 2MASS and IRAC photometry was added. The object IDs are ordered by SOFI \textit{K}-band brightness. The magnitudes of objects located in the SINFONI field of view are listed in Table \ref{tab:magnitudes}.

\section{Cluster surroundings}
\label{sec:colors}

In this section we present the Color Magnitude Diagrams (CMDs) based on SOFI, 2MASS (Figure \ref{fig:IR_colors}) and IRAC (Figure \ref{fig:color_spitzer_both}) photometry to trace stellar properties like temperature, spectral type, reddening and accordingly, the evolutionary stage.

\subsection{SOFI  and 2MASS colors}
The SOFI \textit{J}- and \textit{K}-band photometry\footnote{Note that only the SOFI objects with both \textit{J}- and \textit{K}-band detections are shown since they could be converted to broad-band magnitudes and for this procedure (\textit{J}-- \textit{K}) is required. This has been done to be able to compare the results with the intrinsic colors of ZAMS stars. } for all stars with SINFONI spectra is presented in a (\textit{J}-\textit{K}, \textit{K}) CMD on Figure \ref{fig:color_SOFI_jkk_match}. The magnitudes and colors of main sequence dwarf stars (\citealt{2000Blumetal}; \citealt{1988BesselBrett}) are plotted with an adopted distance of 830 $\pm$ 50 pc. It is readily visible that for most objects the \textit{J}-\textit{K} colors are redder than main-sequence \textit{J}-\textit{K} colors. Assuming that this is caused by extinction, the average reddening is A$_{\rm V}$$\approx$10-20. \textit{J}-\textit{K} will increase for both disk emission and extinction, so disentangling these effects is still a problem, even when the intrinsic stellar color (i.e. spectral type) is known.

The 2MASS \textit{J}-\textit{K}, \textit{H-}\textit{K} color-color diagram  for all stars with SINFONI spectra is shown in Figure \ref{fig:color_2MASS_hkjh_match}. The average extinction is again A$_{\rm V}$$\approx$10-20. From this diagram we can attempt to discern extinction from IR excess. Most objects are aligned close to the dashed extinction line which shows that the bulk of the emission in the \textit{J}-, \textit{H}- and \textit{K}-band originate from their stellar photospheres. Additionally, the fact that we see absorption lines in their \textit{H-} and \textit{K-}band spectra also suggests that the disk contribution is low (See section \ref{sec:spectroscopy}). All but objects 17 and 19 do have significant IR excess in their IRAC bands indicating that there is no clear evidence of disk emission until $\lambda$$>$2.5 $\mu$m for these objects.

\subsection{Extinction map}

A large extinction map of the cluster surroundings as shown in Figure \ref{fig:morphology_dss_IR_spitzer} has been constructed by a combination of 2MASS and SOFI photometry. This map shows the extinction in visual magnitudes and has a resolution of 1.5$\arcmin$. The extinction mapping method is explained in \citet{2001LombardiAlves}. 

There are two extinction blobs in the region, one is centered on the cluster region covered by SINFONI, the other extinction blob falls $\sim$ 4$\arcmin$ north east of that. We can see that star formation in the northern blob (if any) is not as evolved as in the southern blob. Most of the class 0/I sources seem to be concentrated in the southern extinction blob where also the extinction is relatively high. The young high mass stars are also found in this region. The class II objects are spread out over a larger region. This result is in agreement with the IRAC study of \citet{2009Gutermuth}. An average extinction of A$_{\rm V}$$\sim$12 mag in the SINFONI field as derived from the extinction map seems to be consistent with the average extinction found in the previous section (Figures \ref{fig:color_SOFI_jkk_match} and \ref{fig:color_2MASS_hkjh_match}).

\subsection{Spitzer colors}
\label{sec:spitzer}

Figure \ref{fig:color_spitzer_both} presents the color-color diagrams of the IRAC detections in the SINFONI-field. Note that not all objects detected in the \textit{J}-, \textit{H}- and \textit{K}-band are detected in all IRAC bands and vice versa. The colors of photospheres are located near (0,0). Figure \ref{fig:color_SPITZER_3bands_match} shows the (3.6 - 4.5, 4.5 - 5.8) $\mu$m color-color diagram from which we can derive a lower limit of 20 stars (about half of the objects detected by SINFONI) having emission at these wavelengths which cannot be ascribed to reddened photospheres. This suggests that these stars have non-photospheric IR-emission caused by CSM. The other half of the cluster objects lack one or more detections in the first three IRAC bands. Using all four IRAC bands, theoretical \citep{2004Allenetal} and observational (\citealt{2004Megeathetal}; \citealt{2009Gutermuth}) studies found that the IRAC colors of PMS stars trace their evolutionary stage. We apply these color criteria to determine the evolutionary class \citep{1987Lada} of the objects with detections in all IRAC bands. The results for objects in the SINFONI field are shown in the (3.6 - 4.5, 5.8 - 8.0) $\mu$m color-color diagram in Figure \ref{fig:color_spitzer}. Many objects have an IR excess: seven objects are marked as class 0/I and five objects as class II objects.  Furthermore we studied the spatial distribution of the class 0/I and class II sources over a larger region, as shown in Figure \ref{fig:morphology_dss_IR_spitzer}. The class 0/I objects and the younger objects in the SINFONI field (the H~{\sc{ii}}  region, the massive YSO and the Herbig Be star) are more centrally located  as compared to the more evolved class II objects. 

The presence of objects too embedded for SINFONI might be revealed at longer wavelengths by IRAC detections with matching positions in two or more bands. This seems to be the case for 8 objects in the SINFONI field. An interesting result is the detection of the H~{\sc ii}  region VLA 1 (see Figure \ref{fig:morphology_IR_spitzer_color}) at 3.6, 4.5, 5.8 and 8.0 micron with fluxes of respectively 29.1$\pm$1.2 mJy, 61.1$\pm$3.0 mJy, 0.51$\pm$0.043 Jy and 1.01$\pm$0.09 Jy. We note that the detection of the 5.8 $\mu$m band has an offset of $\sim$3-4 $\arcsec$ to the southwest compared to the central location of the VLA 1 position. However, all fluxes are consistent with the SED of VLA 1 \citep{2003PersiTapia}. The IRAC colors of VLA 1 in Figure \ref{fig:color_spitzer} are consistent with typical UCH~{\sc II} colors (e.g. \citealt{2009Fuenteetal}). The H~{\sc ii} region seems extended in the IRAC bands making it difficult to determine the central pixel. Furthermore, on the edges of the dark filament  (in the south east region, see Figure \ref{fig:morphology_IR_spitzer_color}) we observe five locations with detections in all IRAC bands, suggesting that even more deeply embedded young objects are hidden in this featureless region.

The conclusions we draw from the inventory of the photometric data is that for a large fraction of objects (a) highly reddened \textit{J}-, \textit{H}- and \textit{K}-magnitudes ($\sim$10-20 A$_{\rm V}$) are observed and (b) IR excess is present in the IRAC bands. Remarkable are the objects (3, 9, 13 and 17) in Figure \ref{fig:color_2MASS_hkjh_match} showing reddened J, H and K \emph{photospheric} colors while they have IR excess in the IRAC bands. This suggests that emission from the CSM sets in at longer wavelengths than we usually observe for intermediate mass PMS stars \citep{2004AckeAncker}.

\section{SINFONI near-infrared spectroscopy}
\label{sec:spectroscopy}
We use SINFONI \textit{H}- and \textit{K}-band spectra \footnote{SINFONI spectra are available in electronic form upon request: kmaaskan@science.uva.nl} to classify the stars by comparing absorption and emission lines (intensity or line strength ratios) with reference spectra of  \citet{2009Rayneretal} and \citet{2005Cushingetal}. This provides a powerful tool for examining a variety of stellar properties such as spectral type, mass and the presence of CSM.  A summary of the stellar properties of all objects detected by SINFONI is presented in Table \ref{tab:dataoverview}.  In the next sections we study the spectroscopic diagnostics for early- and late-type stars.

\begin{figure}[]
\includegraphics[scale= 0.7]{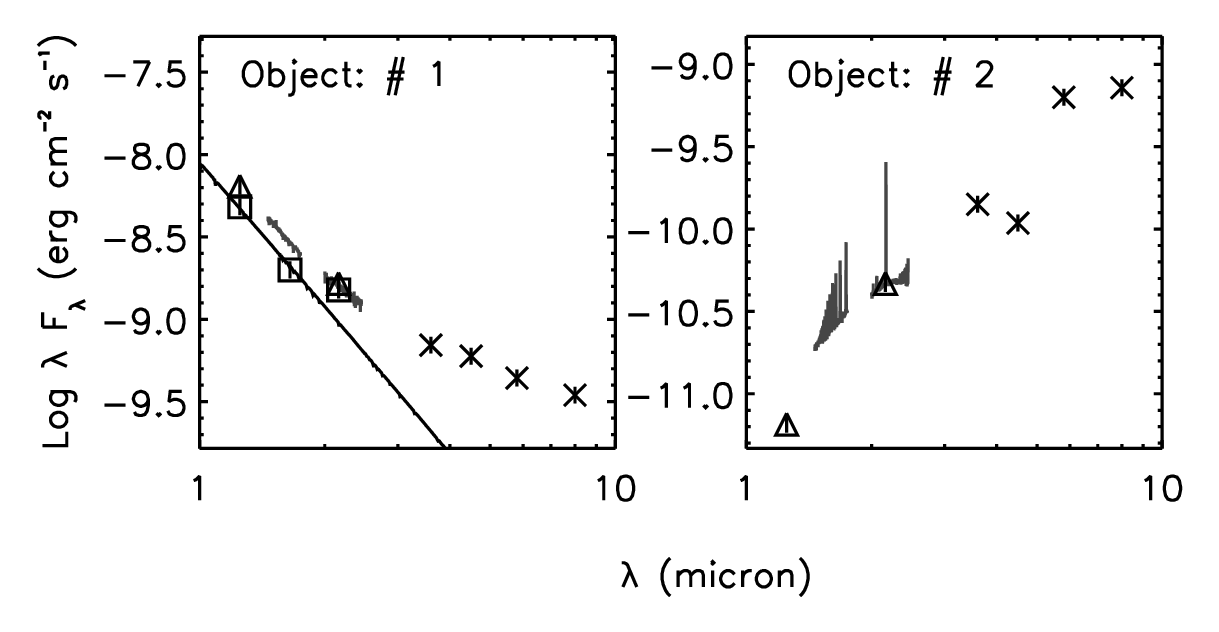}
\label{fig:disk_mcmax_object3}

\caption{\label{fig:sed_object_1_2}The spectral energy distribution of object 1 (left panel) and 2 (right panel). The Kurucz model for a B0 star is shown for object 1. The de-reddened slope of the (photospheric) SINFONI spectra provides an additional consistency check for the classification method and the derived extinction. The SOFI, 2MASS and IRAC detections are marked with triangles, squares and crosses respectively. }
\end{figure}

\begin{figure*}
\centering
\includegraphics[width=\textwidth]{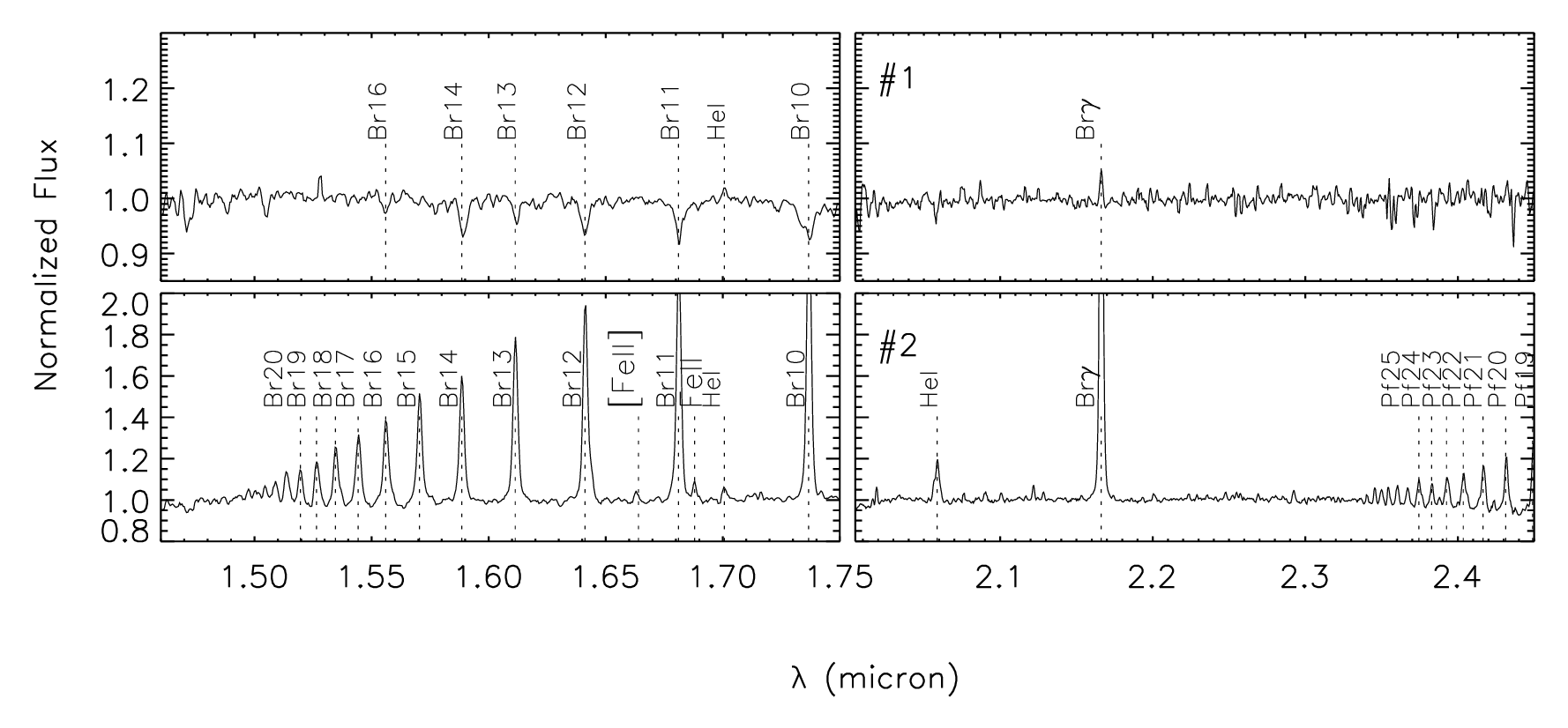} 

\caption{\label{fig:high_mass_spectra}Normalized SINFONI \textit{H}- and \textit{K}-band spectra of the two brightest stars detected in IRAS 06084-0611. The spectrum of object 1 is characterized by absorption lines of the hydrogen Brackett series. 
The spectrum of object 2 shows numerous hydrogen lines in emission of which Br${\gamma}$ (2.166 $\mu$m) is the most prominent. The spectra of the other objects in the field are shown in Figure \ref{fig:allspectra_1_21} and \ref{fig:allspectra_22_39}. }
\end{figure*}

\subsection{Early-type stars}
There are three early-type objects in the field: a B0.5 star embedded in an H~{\sc ii}  region (VLA 1, \citealt{2002Gomezetal}) with an IR counterpart detected in the IRAC bands, a deeply embedded early-B type (object 1) and a Herbig Be star identified in this paper as object 2. Their basic properties are shown in Table \ref{tab:dataoverview}. We proceed with a discussion of the properties of object 1 and 2.

\emph{Object 1} is a bright and highly reddened massive YSO with IR excess in the IRAC bands (see the SED in the left panel of Figure \ref{fig:sed_object_1_2}). Figure \ref{fig:high_mass_spectra} shows the SINFONI spectra in the \textit{H}- and \textit{K}-band. The \textit{H}-band spectrum shows a photospheric absorption pattern in the Brackett series. The presence of these lines are typical for B-type stars (\citealt{1997Blumetal}; \citealt{1998Hansonetal}). However, the Brackett absorption lines are unexpected: the Br-15 is not visible at all and Br-14 is wider and deeper than Br-13. This suggests that these lines are contaminated by other emission or absorption features, though their origin is as yet unclear. Furthermore we report weak He~{\sc i} (1.70 $\mu$m) in emission. 
Remarkable in the \textit{H-}band spectra are two absorption lines at 1.490 and 1.505 $\mu$m. These lines can be identified as Mg~{\sc i}  absorption \citep{2009Rayneretal}; however, note that neutral magnesium is not expected to originate from the photosphere of a massive star since the stellar temperature is too high. A strong telluric emission line is present at 1.506 $\mu$m, but at 1.490 $\mu$m the telluric contamination is minimal. The identification and origin of these absorption lines demand further exploration and a study of follow-up VLT/X-Shooter observations is planned. 
In the \textit{K}-band, weak and narrow Br-$\gamma$ emission (2.166 $\mu$m) is visible. Br-$\gamma$ emission could be produced by e.g. a disk wind caused by the interaction between the UV photons and the ionized upper layer of the disk \citep{2006Biketal}, magnetospheric accretion \citep{2005Ancker} or inner disk accretion \citep{2004Muzerolleetal}. Furthermore, the emission could be of nebular origin, as suggested by the relative small line width, though the quality of our data is insufficient to quantify these scenarios further and a comprehensive understanding has yet to emerge.

We estimate a spectral type using the classification scheme from \citet{1996Hansonetal,1998Hansonetal} for the \textit{K}- and \textit{H}-band spectra, respectively. An EW of $\sim$2 $\AA$ is measured for the Br-11 absorption line (1.6814 $\mu$m). Comparing this to the line strength as a function of spectral type, object 1 is consistent with a spectral type of B0 ($\pm$ 2). The 1.700 He~{\sc i} $\mu$m line is visible in emission, therefore we cannot use that line for further classification. Note that veiling, most likely present, would result in assigning a later spectral type. No other absorption or emission line features could be used for spectral classification. We derive an extinction of A$_{\rm V}$=21.0 mag using the 2MASS \textit{J}-\textit{H} color which is relatively independent of spectral type in the Rayleigh-Jeans domain for spectral types earlier than F.  Another estimate of the spectral type can be derived by assuming that this object is a ZAMS cluster member located at a distance of 830 $\pm$ 50 pc; we find an absolute magnitude M$_{\rm K}$ of -2.44 mag, typical for an early-B ZAMS star \citep{2000Blumetal}.

\emph{Object 2} shows prominent hydrogen Brackett and Pfund lines and the helium line at 2.0581 $\mu$m in emission (Figure \ref{fig:high_mass_spectra}). The flux in the \textit{H}- and \textit{K}-band is dominated by CSM which is irradiated by the central object and hence produces the abundant emission lines. We identify Fe~{\sc ii} (1.688 $\mu$m) and [Fe~{\sc ii}] (1.664 $\mu$m) emission lines. Strong PAH features at 8.6, 11.2 and 12.7 $\mu$m are present (\citealt{2003PersiTapia}; \citealt{2009Boersmaetal}). The peculiar shape of the SED at IRAC wavelengths (see the right panel on Figure \ref{fig:sed_object_1_2}) might be explained by such strong dust features. Object 2 has a radio counterpart (VLA 4) which shows time variability and has a negative spectral index. \cite{2003PersiTapia} studied the IR energy distribution (assuming a distance of 1 kpc) and obtained a luminosity consistent with the presence of a very young star somewhat later than B3. However, the He~{\sc i} 1.70 $\mu$m recombination emission line suggests a spectral type earlier than B2.5V \citep{2000ClarkSteele}.

\subsection{Late-type stars}
\label{sec:stellar_content}
\begin{figure*}
\label{sec:appspectra}

\centering
\includegraphics[width=0.7\textwidth]{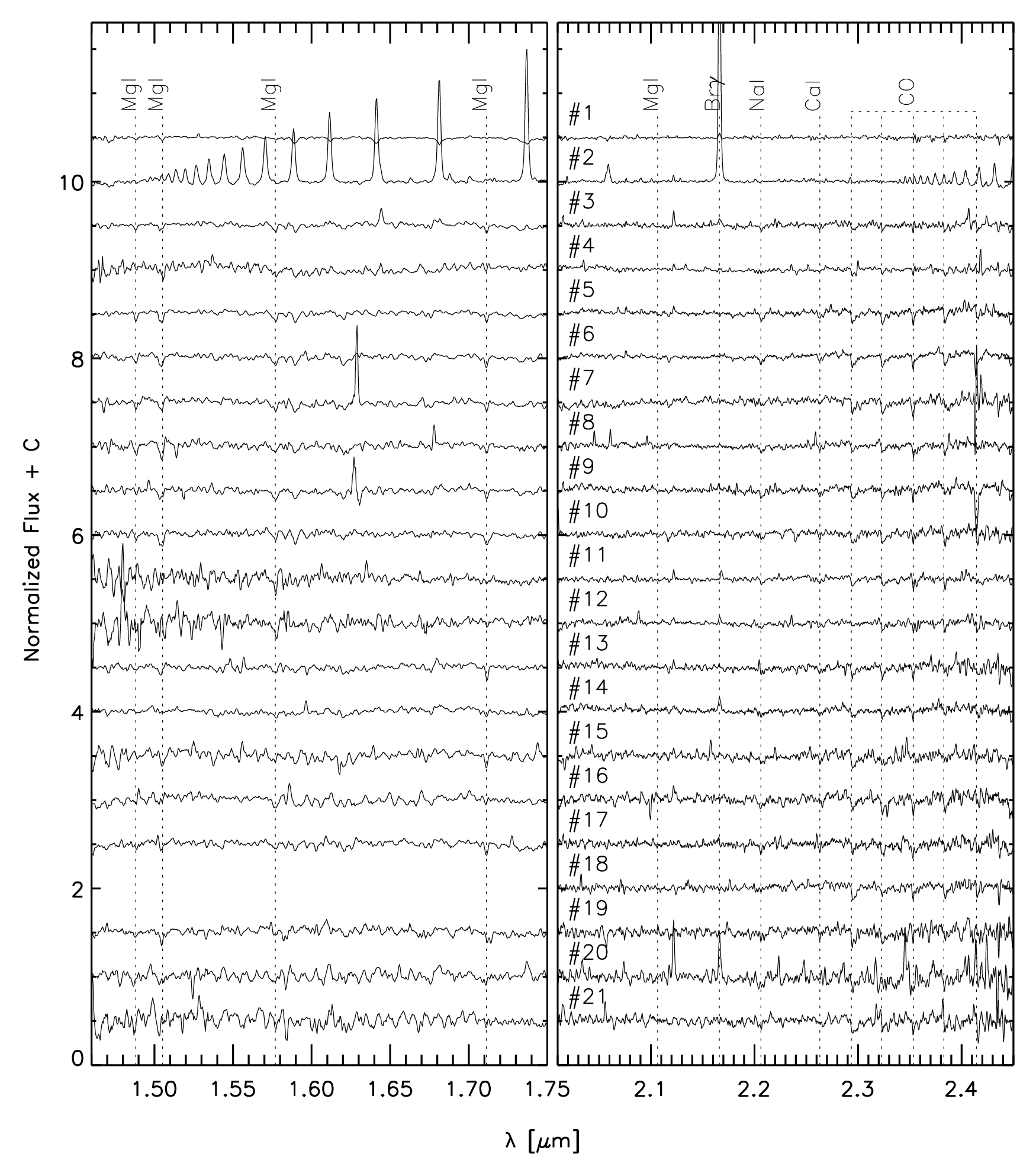} 
\caption{\label{fig:allspectra_1_21} Normalized SINFONI \textit{H}-band (left) and \textit{K}-band (right) spectra of the first 21 objects. For most of these objects we were able to derive the spectral type (Table \ref{tab:dataoverview}). The location of some of the absorption lines used for the classification of the stars are shown by the vertical dashed lines. The combination of Mg~{\sc i} , Na~{\sc i}  and Ca~{\sc i}  absorption lines proved to be the best temperature tracers. The width of the CO lines in the \textit{K}-band gives an indication of the luminosity class. }
\end{figure*}

\begin{figure*}
\centering
\includegraphics[width=0.7\textwidth]{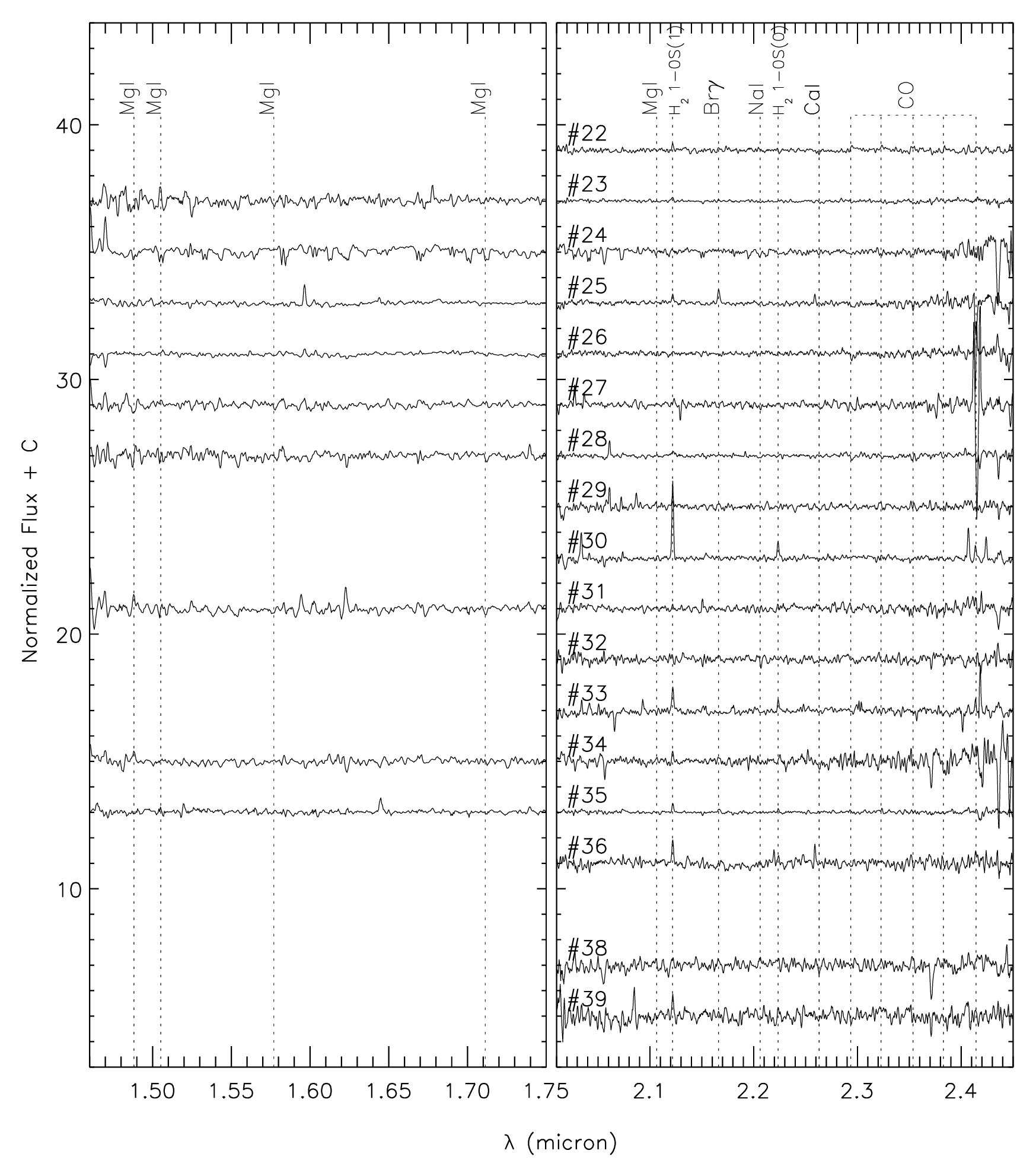} 
\caption{\label{fig:allspectra_22_39} 
Normalized SINFONI \textit{H}-band (left) and \textit{K}-band (right) spectra of the remaining objects. The spectra with very bad S/N are not shown.  }
\end{figure*}

For the spectral classification of low-mass stars, the SINFONI spectra were visually compared with reference spectra from the spectral library of \citet{2009Rayneretal} and \citet{2005Cushingetal}. In this way, the spectral types of 14 stars were determined in a range from early G to mid-K with generally an uncertainty of $\sim$1-2 in subclass. Spectra dominated by CSM or too low S/N could not be classified. Figures \ref{fig:allspectra_1_21} and \ref{fig:allspectra_22_39} show the SINFONI \textit{H}- and \textit{K}-band spectra for all the objects in the field. Table \ref{tab:dataoverview} shows the results of the spectral classification for these objects. 

The temperature of the classified objects is derived using the spectral type to temperature relation of \citet{1995KenyonHartmann}. It has been found by \citet{1979CohenKuhi} that the temperature of PMS stars might be overestimated and translate to non-systematic errors, resulting in an error of about 500 K for G stars and about 200 K for late K-type stars. This uncertainty introduces the dominant source of error. 

Important absorption lines for the temperature determination of low-mass stars are neutral metal absorption features. Mg~{\sc i}  (1.50, 1.53 and 1.71 $\mu$m) shows a gradual variation as a function of spectral type: these lines are strongest in K and early M dwarfs. The Ca~{\sc i}  (2.26 $\mu$m) triplet absorption line is present in early M, K and G stars and weakens in the late-M spectral types. The Na~{\sc i}  doublet (2.20 $\mu$m) becomes visible in early G stars and is strongest in mid- to late-type M dwarfs. CO overtone bands in the \textit{K}-band ($\sim$2.29 - 2.5 $\mu$m) are strongest in supergiants and become progressively weaker with decreasing luminosity. Therefore, they provide good surface gravity indicators since PMS stars have lower surface gravities while evolving to the main sequence. CO overtone bands can also be observed in emission. In that case they are most likely originating from a circumstellar disk in combination with a stellar wind (e.g. \citealt{1995Chandleretal,2004BikThi, 2010Wheelwrightetal}). This is observed in the YSO object 22 (Figure \ref{fig:allspectra_22_39}).   

Visual extinctions have been derived by comparison of SOFI and 2MASS colors to the intrinsic giant and dwarf colors of \citet{1983Koornneef2} and using the extinction law of \citet{1989Cardellietal} to be able to express the extinction in $A_V$. 
For spectra with luminosity class IV, the extinction has been derived from the average of the giant and dwarf colors of \citet{1983Koornneef2}. 

A representative Kurucz model and de-reddened SINFONI spectra have been over-plotted for objects with known spectral type (e.g. for object 1 in Figure \ref{fig:sed_object_1_2}). The slope of the de-reddened SINFONI spectrum provides an additional consistency check for the classification method: the spectral type is consistent with the derived extinction if the de-reddened photospheric SINFONI spectrum fits the Kurucz model and the de-reddened photometry. Most de-reddened objects show a good fit which implies that the extinction has been correctly estimated.

\begin{table*}

\caption{\normalsize{\textsc{properties of objects located in the SINFONI field of view}}}

\scriptsize{
\hspace{-0.5cm}\begin{tabular}{ c  c c l c c c c c c c c c c c}
\multicolumn{15}{l}{ }\\
\hline 
\hline
\multicolumn{1}{c}{\textit{ } } &\multicolumn{2}{c}{\textit{(J2000)} }  & \multicolumn{4}{c}{\textit{Available photometric data} }&\multicolumn{7}{c}{\textit{Objects properties} }&\multicolumn{1}{c}{cluster }     \\
Object \#  &  RA (h m s) &   Dec ($^\circ$ $\arcmin$ $\arcsec$)  &   SOFI & 2MASS  &  IRAC    &   Radio&  Sp. Type$^a$  &  Lum$^a$  &  IR$_{\rm ex}$$^b$  &  class$^c$ & T$_{\rm eff}^d$  & A$_V^e$ & R$_{\odot}^f$ &member\\
\hline

\input{tab2_16743.tex}
\hline
\multicolumn{14}{l}{ }\\
\multicolumn{14}{l}{$^a$: Based on spectral comparison with reference spectra from \cite{2009Rayneretal}.}\\
\multicolumn{14}{l}{$^b$: IR excess based on 2MASS, SOFI and IRAC photometry compared to Kurucz models \citep{1993Kurucz} and IRAC color criteria as described in section \ref{sec:spitzer}.}\\
\multicolumn{14}{l}{$^c$: Evolutionary class \citep{1987Lada} derived from the IRAC colors as defined by \citet{2004Megeathetal}.  }\\
\multicolumn{14}{l}{$^d$: Temperature from \citet{1995KenyonHartmann}, averaged between the spectral type range.}\\
\multicolumn{14}{l}{$^e$: A$_{\rm V}$ derived using intrinsic colors of \cite{1983Koornneef2}.}\\
\multicolumn{14}{l}{$^f$: Result of fitting with Kurucz  \citep{1993Kurucz} model. A distance of 830 $\pm$ 50 pc is assumed \citep{1976HerbstRacine}.}\\
\multicolumn{14}{l}{$^\dagger$:  Based on radio continuum flux  \citet{2002Gomezetal} estimate the ionizing object to be of spectral type B0.5. }   \\
\multicolumn{14}{l}{$^{\dagger\dagger}$: Based on the infrared luminosity  \citet{2003PersiTapia} estimate the presence of a very young star somewhat later than B3.}\\

\end{tabular} 
}
\label{tab:dataoverview}

\end{table*}

\subsection{Cluster membership}
To confirm the cluster membership of the objects, we looked for the following characteristics: (a) highly reddened photometry; (b) IR-excess in the IRAC bands; (c) evidence of jets/ouflows by H$_2$ and [Fe~{\sc ii}] emission associated with the object and (d) the presence of a radio (VLA) counterpart. In this way we identified 28 objects as cluster members. We could not confirm the cluster membership of 10 objects. Object 14 is excluded as a cluster member, since the spectral type is too late for its brightness. It is therefore marked as a foreground star.

\subsection{Jets and outflows}

We use the green H$_2$ and blue [Fe~{\sc ii}] line maps (Figure \ref{fig:subSINFONI}) to search for emission blobs in the field that trace star formation activity. H$_2$ emission can be related to outflows since thermal H$_2$ emission in the near-IR can originate from shocked neutral gas being heated up to a few 1000 K. Alternatively, non-thermal emission can be caused by fluorescence of non-ionizing UV photons. These mechanisms can be distinguished as they populate different energy levels and thus produce different line ratios (see \citet{2010Biketal,2011Wangetal} for discussions of these mechanisms for RCW 34 and the S255 complex).  For the three northern H$_2$ emission blobs in IRAS 06084-0611, the outflow nature has been confirmed making use of hydrogen emission line ratios \citep{2008Pugaetal}. The morphology of the emitting regions (compact blobs) would suggest an outflow nature, the fluorescence emission is usually diffuse and extended. Another tracer of active star formation is [Fe~{\sc ii}] emission, often observed in star forming regions associated with shocked gas from jets and outflows or photon dominated regions.   

H$_2$ and a few [Fe~{\sc ii}] emission regions are spread out over the cluster (Figure \ref{fig:subSINFONI}). H$_2$ emission blobs are found near object 3, 4 and 30 and might they trace outflows from these objects. For the other half we find emission regions lacking central objects; they could be too deeply embedded or are Photon Dominated Regions (PDRs) excited by the UV radiation from these objects. 

Our observations identify object 30 as the IR counterpart of the thermal non-variable radio source VLA 7, the suggested central object powering the massive CO outflow. Recall that VLA 7 is the suggested central object powering the massive CO outflow that extends $\sim6\arcmin$  from SE to NW (section \ref{intro}). The SINFONI \textit{K}-band spectrum of this object shows strong H$_2$ emission lines (see Figure \ref{fig:allspectra_22_39}) and the IRAC detection in the first three bands shows highly reddened colors, suggesting the presence of a protostar.

\section{Stellar population: content and age}

In this section we discuss the global properties of the stellar content in more detail. We present the Hertzsprung-Russell (HR) diagram and compare the observed parameters with PMS evolutionary tracks. 

\label{sec:HerRus}

\begin{figure}[t]

\subfloat[]{
\includegraphics[scale= 0.6]{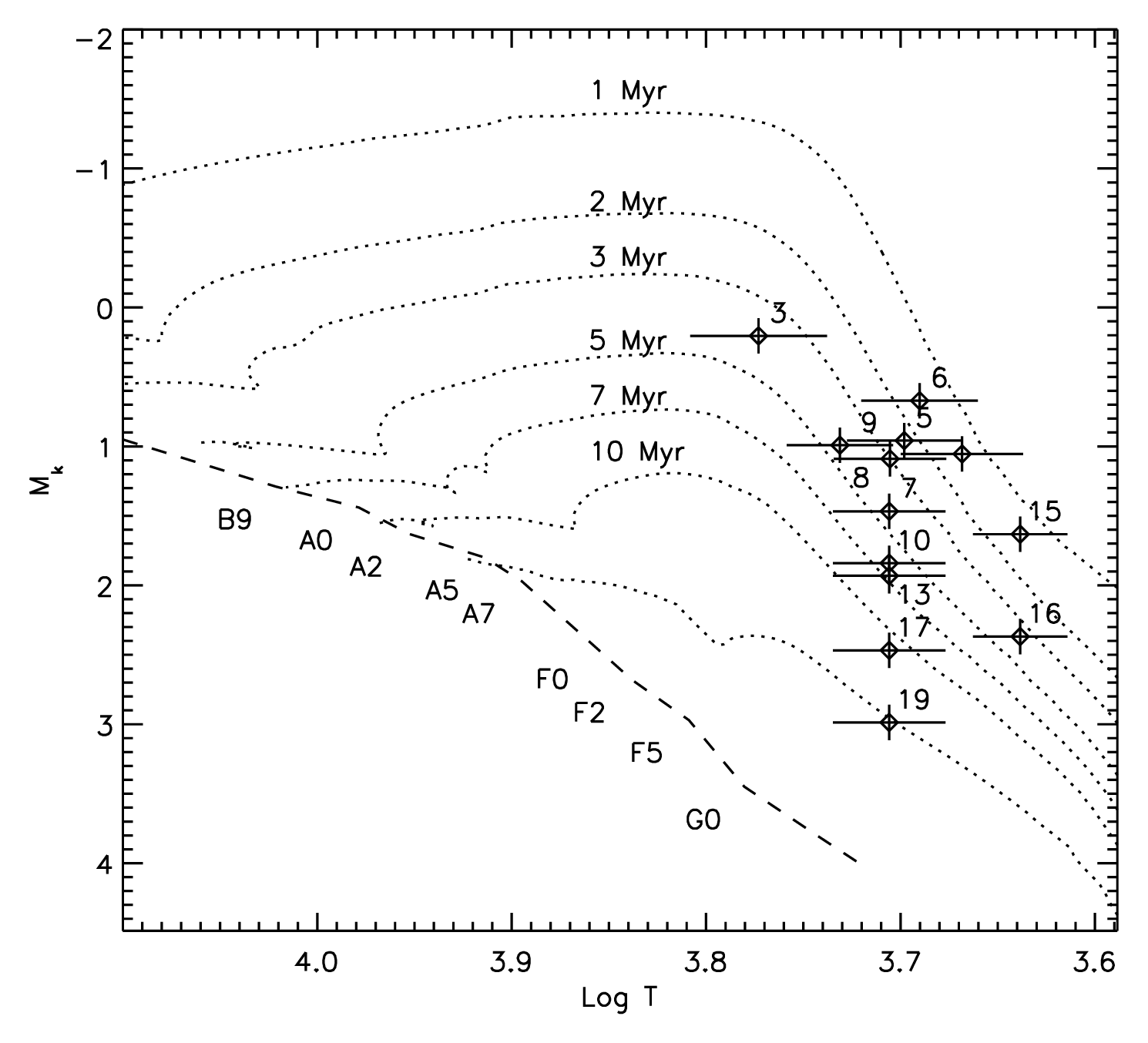}
\label{fig:HR_age}
}
\\ \subfloat[]{
\includegraphics[scale= 0.6]{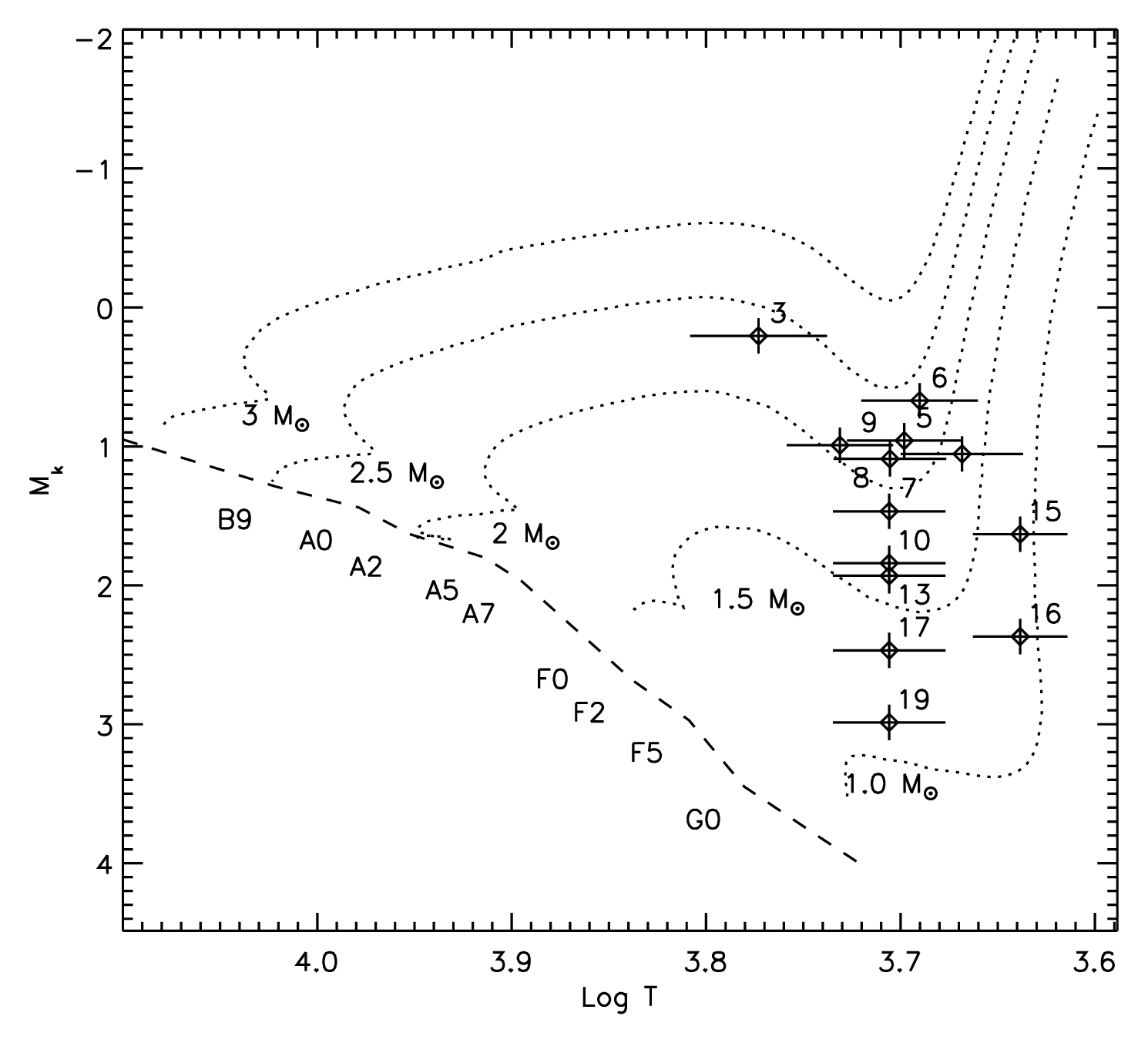}
\label{fig:HR_mass}
}
\caption{\label{fig:HR}HR diagram of objects classified by their SINFONI spectra.  \textit{Top:} the extinction corrected M$_{\rm K}$ vs. log(T) HR diagram. The \textit{K}-band magnitude has been corrected for a distance of 830 $\pm$ 50 pc. Over plotted with a dashed line is the ZAMS isochrone from \citet{2000Blumetal} and the dotted lines represent the isochrones from \citet{2009DaRioetal} between 1 and 10 Myr calculated from the \citet{2000Siessetal} evolutionary models. The bottom panel shows the same diagram with overplotted the pre-main-sequence evolutionary tracks for 3, 2.5, 2, 1.5 and 1 M$_{\odot}$ stars from \citet{2009DaRioetal}.}
\end{figure}

The HR diagram of IRAS 06084-0611 is shown in Figure \ref{fig:HR}. Over-plotted are the ZAMS-isochrone from \citet{2000Blumetal}, isochrones (upper panel) between 0.1 and 10 Myr and PMS evolutionary tracks (lower panel) from 1 up to 5 M$_{\odot}$. The latter isochrones have been taken from  \citet{2009DaRioetal} calculated from the \citet{2000Siessetal} evolutionary models. An estimate of the average age of the cluster provides more insight into the cluster history.  Furthermore, comparison of the positions in the HR diagram to the evolutionary tracks provides information on the mass of the classified stars. 

For the derivation of the absolute magnitude, a distance of 830 $\pm$ 50 pc \citep{1976HerbstRacine} was adopted and the objects have been corrected for foreground extinction. For all objects, the uncertainty of the absolute magnitude has been derived taking into account four effects: the first and second are the photometric and distance uncertainty. The third is a consequence of the uncertainty in extinction which is a reflection of the uncertainty of the spectral type and the variation of reddening between the different photometric colors (see Table \ref{tab:dataoverview}). The fourth is the uncertainty in distance of the cluster. For all objects, the first three sources of uncertainty are typically a few percent while the latter translates to a dominant source of uncertainty of $\pm$0.13 mag.

The assumption that all the objects have the same distance might be incorrect if the cluster is extended along the line of sight (i.e. if the group of younger class 0/I objects and the group of class II sources have different distances). In this study we assume the stars to be equidistant.

We find that most objects lie on the evolutionary tracks of $\sim$1.5-2.5 M$_{\odot}$ stars (Figure \ref{fig:HR_mass})  and thus the majority are intermediate-mass PMS stars. The age of the stars ranges between 1 and 10 Myrs (Figure \ref{fig:HR_age}) and we find a median age of $\sim$4 Myr. This result can be considered as an upper limit of the cluster age since no ZAMS intermediate-mass stars have been detected. However, the spread in age is large and suggests that the stars might not all originate from the same cluster or have a complex formation history. We will discuss this scenario further in the discussion section.

We note that the number of objects with known spectral type is less than half of the objects observed in the SOFI \textit{K}-band. This implies that there is a selection effect which excludes objects that do not show photospheric absorption lines in the \textit{H}- and \textit{K}-band. This is the case for objects 4, 11, 12, 18 and 20 and objects with low signal to noise ratio spectra. Since the detection of circumstellar emission is an indication that an object is a young PMS star, this would lower the average cluster age. The fainter objects might include young objects still deeply embedded in the parental cloud, which obviously lowers the cluster age estimate as well. So while the group of stars shown in the HR diagram is $\sim$4 Myrs old, the group of stars which could not be classified might be (much) younger.  We can relate this result with our findings from the IRAC colors (section \ref{sec:spitzer}) which show that the class II objects are spread out over a larger region as compared with the class 0/I objects. All the class II objects are identified in the HR diagram. Six out of seven class 0/I objects could not be classified.  It seems plausible that the age derived from the HR diagram is primarily dominated by the class II objects.

\section{Discussion}
\label{sec:discussion}

We start with a review of the observed stellar properties of the cluster. We can distinguish two groups based on their age and position. 

The first group of objects include the massive stars (object 1, 2 and the H~{\sc ii}  region) and the class 0/I objects. CSM around object 1 and 2 and the presence of the H~{\sc ii} region suggest that these objects belong to a very young population of stars less than a million years old. Since the massive objects are found in the center of the class 0/I objects (as demonstrated by their spatial distribution on the IRAC field in Figure \ref{fig:morphology_dss_IR_spitzer}), this suggest that most class 0/I objects might be associated with this younger population. For several of the class 0/I objects we have SINFONI spectra (object 4, 18, 22, 29). Their spectra do not show photospheric absorption lines. They have significant excess in the \textit{H-}, \textit{K-} and IRAC-bands but lack detections in the \textit{J}-band, suggesting that these objects are strongly reddened and surrounded by CSM. Furthermore we observe several H$_2$ blobs lacking central objects in our SINFONI data, indicating early star formation activities deeply embedded in the cluster.

The second group consists of a population of class II objects which appear more diffused outward in the IRAC field (Figure \ref{fig:morphology_dss_IR_spitzer}) and have typical ages of a few Myrs. The class II objects in the SINFONI field can be associated with the spectroscopically classified stars as most of them (with detections in all IRAC bands) are class II objects (see Table \ref{tab:dataoverview}). We classified 13 intermediate-mass PMS stars, and found that they are typically of G-K spectral type with luminosity class IV and will evolve to stars with late B, A or early F spectral type. With an extinction of $\sim$10-20 A$_{\rm V}$ mag, they show IR excess in the IRAC bands starting at wavelengths $>$2 $\mu$m. These objects are placed in an HR diagram (Figure \ref{fig:HR}), and comparison with evolutionary tracks shows that they represent a group of stars with a median age of $\sim$4 Myr. Note that no massive stars are observed in this group.

We will now attempt to reconstruct the star formation history of this region and examine how they are related in age and location. IRAS 06084-0611 is a cluster spanning a wider range of age than is typical of stars formed in any individual star-forming region. The presence of young massive objects (group 1) suggests that a part of the cluster is significantly younger than the stars belonging to the $\sim$4 Myr population (group 2). Though we cannot prove a possible physical relation from our dataset, it is reasonable to suggest that despite the range in age, the star formation history of group 1 and 2 in the cluster might be closely related. However, what we really observe is sequential star formation: the most massive stars of group 1 have formed sequential to the initial cluster of intermediate mass PMS stars (group 2).

There is very strong differential extinction in the SINFONI field, particularly towards the southeast. The extended H$_2$ emission in the lower half of the SINFONI field (Figure \ref{fig:subSINFONI}) could be originating from a lower density PDR where the filament is less dense. This might also explain why object 4, 11, 12, 21 and 23 are not detected in the \textit{J-}band.  On a larger scale, the IRAC color image in Figure \ref{fig:morphology_IR_spitzer_color}  and the extinction map in Figure \ref{fig:morphology_dss_IR_spitzer} show dark filamentary structures indicating large extinction. On the borders of the dark filament in the SINFONI field we find 5 objects as well as the H~{\sc ii} region (VLA1) only detected in the IRAC bands. This population, deeply embedded in optically thick molecular clouds, is illustrative for the youth of this cluster region. 

The SINFONI field is too small to observe gradients in the distribution of massive stars and class 0/I objects and the intermediate-mass PMS stars  (See Figure \ref{fig:objects_VLA}). However, in the IRAC field (Figure \ref{fig:morphology_dss_IR_spitzer}), the spatial distribution of class 0/I and class II objects shows that the class 0/I objects are concentrated at shorter distances around the massive objects (object 1, 2 and the H~{\sc ii} region) as compared to the class II objects. This suggests that the intermediate mass cluster (group 2) could have lost enough natal cloud mass in the first million years since formation that the stars are no longer gravitationally bound. They are in the process of diffusing outward and hence appear more spread out compared to the younger class 0/I objects. Taking into account that the SINFONI observations show many infant star formation features like outflows and strong reddened colors, and that class 0/I objects are younger than class II objects, it seems that the center of IRAS 06084-0611 contains a relatively higher fraction of young objects than its surroundings. A possible scenario for this observation would be sequential star formation along the line of sight;  in front or behind the initial cluster from our perspective. 

The observation that the most massive stars are concentrated towards the center of the cluster is observed in many more young clusters \citep{1998HillenbrandHartmann,2004Gouliermisetal}. Mass segregation to the center of the cluster could be primordial or dynamical: the most massive stars can be concentrated at or near the center or migrate into the center of the cluster due to two-body interactions. \citet{2009Allisonetal} describe an ensemble of simulations of cool, clumpy (fractal) clusters and they show that during the early life of such a cluster often mass segregation takes place on timescales far shorter than expected from simple models. The results of these N-body simulations look quite similar to the current situation in IRAS 06084-0611.

Our findings show an interesting similarity to those found in the S255 complex \citep{2010Biketal,2011Wangetal}. Also in this regions, a group of intermediate mass stars (up to a few Myrs) seems to have formed after star formation in a smaller region containing younger, more massive stars. As for IRAS 06084-0611, sequential star formation might be at work too.

\section{Conclusions}
The results of our study can be summarized as follows:

\begin{enumerate}

\item We detect two high mass stars in our SINFONI data: a massive YSO consistent with an early-B spectral type and a Herbig Be star. Moreover we detected the IR counterpart of the H~{\sc ii} region VLA1 in the IRAC bands.

\item The spectral types of 13 objects are determined in the star forming region IRAS 06084-0611 using SINFONI \textit{H}- and \textit{K}-band spectra. Most objects for which a spectral type could be obtained are PMS intermediate-mass stars of G - K spectral type. 

\item Using 2MASS, SOFI and IRAC photometry we find that the extinction shows a large variation over the SINFONI field ranging between A$_{\rm V}$$\sim$10-20 mag.

\item We derive a median cluster age of $\sim$4 Myr. This seems to be in contradiction with the presence of embedded class 0/I objects and two massive objects (a massive YSO and an H~{\sc ii}  region) with ages of $\sim$1 Myr.

\item The younger class 0/I objects and the massive objects are located in a smaller region as compared to the class II objects,  with rough scale size radii of $\sim$0.5 pc and  $\sim$1.25 pc respectively. A possible scenario for this observation would be sequential star formation along the line of sight: from the initial formation of a cluster of intermediate-mass stars to the formation of more massive stars.

\end{enumerate}

\begin{acknowledgements}
K. M. thanks the staff of the MPIA for their hospitalities. The authors thank F. Eisenhauer for providing the data reduction software SPRED, R. Davies for providing his routines to remove the OH line residuals, N. Da Rio for providing the isochrones and evolutionary tracks and Jouni Kainulainen for sharing his expertise in making the NIR extinction map. The authors thank the Paranal staff for carrying out the observations. 

This publication makes use of data products from the Two Micron All Sky Survey, which is a joint project of the University of Massachusetts and the Infrared Processing and Analysis Center/California Institute of Technology, funded by the National Aeronautics and Space Administration and the National Science Foundation.
This work is based on observations made with the Spitzer Space Telescope, which is operated by the Jet Propulsion Laboratory, California Institute of Technology under a contract with the National Aeronautics and Space Administration (NASA). 
E.P. is partially funded by the Consolider-Ingenio 2010 Program CSD2006-00070

We thank the anonymous referee for the careful review of the manuscript that helped to improve the paper.

 \end{acknowledgements}

\bibliography{bib_16743.bib} 




\end{document}

%% file: tab1_16743.tex
    VLA1&\dots&\dots&\dots&\dots&  \dots&       9.95$\pm$  0.05 &8.67$\pm$  0.05& 5.91 $\pm$0.10&4.49 $\pm$  0.10\\
      1 &  14.02 $\pm$  0.08 &   9.33 $\pm$  0.01 &  13.61 $\pm$  0.06 &  11.11 $\pm$  0.04 &   9.33 $\pm$  0.02 &   7.56 $\pm$  0.00 &   6.80 $\pm$  0.00 &   6.26 $\pm$  0.00 &   5.52 $\pm$  0.01 \\ 
      2 &  15.06 $\pm$  0.13 &  10.83 $\pm$  0.03 & \dots &  \dots &  \dots &   8.05 $\pm$  0.01 &   7.61 $\pm$  0.02 &   4.94 $\pm$  0.01 &   3.81 $\pm$  0.02 \\ 
      3 &  13.73 $\pm$  0.07 &  10.97 $\pm$  0.03 &  13.60 $\pm$  0.03 &  11.96 $\pm$  0.03 &  10.99 $\pm$  0.02 &   9.63 $\pm$  0.02 &   8.79 $\pm$  0.02 &   8.04 $\pm$  0.06 &   7.12 $\pm$  0.08 \\ 
      4 &  \dots &  10.98 $\pm$  0.03 & \dots &  13.19 $\pm$  0.05 &  10.73 $\pm$  0.05 &   8.61 $\pm$  0.00 &   7.74 $\pm$  0.00 &   7.09 $\pm$  0.01 &   6.59 $\pm$  0.02 \\ 
      5 &  12.79 $\pm$  0.05 &  11.01 $\pm$  0.03 &  12.67 $\pm$  0.04 &  \dots &  \dots &   9.75 $\pm$  0.04 &   9.56 $\pm$  0.08 &  \dots &  \dots \\ 
      6 &  15.10 $\pm$  0.14 &  11.69 $\pm$  0.04 &  14.76 $\pm$  0.05 &  \dots &  11.61 $\pm$  0.05 &   9.48 $\pm$  0.04 &  \dots &  \dots &  \dots \\ 
      7 &  14.34 $\pm$  0.10 &  11.99 $\pm$  0.05 &  14.30 $\pm$  0.04 &  12.78 $\pm$  0.04 &  12.12 $\pm$  0.04 &  11.85 $\pm$  0.06 &  11.89 $\pm$  0.12 &  \dots &  \dots \\ 
      8 &  15.55 $\pm$  0.17 &  12.11 $\pm$  0.05 & \dots &  \dots &  12.17 $\pm$  0.04 &  10.18 $\pm$  0.01 &  \dots &  \dots &  \dots \\ 
      9 &  15.57 $\pm$  0.17 &  12.12 $\pm$  0.05 &  15.39 $\pm$  0.04 &  13.26 $\pm$  0.04 &  12.10 $\pm$  0.02 &  10.72 $\pm$  0.01 &  10.08 $\pm$  0.01 &   9.50 $\pm$  0.03 &   8.60 $\pm$  0.03 \\ 
     10 &  14.59 $\pm$  0.11 &  12.25 $\pm$  0.05 & \dots &  \dots &  12.41 $\pm$  0.06 &  11.38 $\pm$  0.04 &  \dots &  \dots &  \dots \\ 
     11 &  \dots &  12.26 $\pm$  0.05 & \dots &  14.27 $\pm$  0.15 &  11.87 $\pm$  0.06 &   9.69 $\pm$  0.03 &   8.78 $\pm$  0.04 &   7.07 $\pm$  0.06 &  \dots \\ 
     12 &  \dots &  12.40 $\pm$  0.06 & \dots &  13.99 $\pm$  0.11 &  12.17 $\pm$  0.05 &  10.28 $\pm$  0.01 &   9.49 $\pm$  0.01 &   8.55 $\pm$  0.02 &  \dots \\ 
     13 &  14.84 $\pm$  0.12 &  12.46 $\pm$  0.06 &  14.86 $\pm$  0.04 &  13.40 $\pm$  0.03 &  12.74 $\pm$  0.03 &  12.00 $\pm$  0.02 &  11.70 $\pm$  0.04 &  11.40 $\pm$  0.09 &  10.49 $\pm$  0.05 \\ 
     14 &  14.62 $\pm$  0.11 &  12.50 $\pm$  0.06 & \dots &  \dots &  \dots &  \dots &  \dots &  \dots &  \dots \\ 
     15 &  16.11 $\pm$  0.22 &  12.76 $\pm$  0.07 & \dots&  14.00 $\pm$  0.03 &  12.95 $\pm$  0.03 &  11.39 $\pm$  0.02 &  10.74 $\pm$  0.02 &  10.21 $\pm$  0.06 &   9.45 $\pm$  0.05 \\ 
     16 &  15.49 $\pm$  0.17 &  12.79 $\pm$  0.07 & \dots &  \dots &  \dots &  \dots &  \dots &  \dots &  \dots \\ 
     17 &  14.89 $\pm$  0.12 &  12.87 $\pm$  0.07 &  15.32 $\pm$  0.04 &  13.84 $\pm$  0.03 &  13.18 $\pm$  0.04 &  11.83 $\pm$  0.04 &  11.44 $\pm$  0.06 &  10.62 $\pm$  0.10 &   9.58 $\pm$  0.08 \\ 
     18 &  \dots &  13.24 $\pm$  0.00 & \dots & \dots &  13.01 $\pm$  0.04 &  10.84 $\pm$  0.01 &   9.91 $\pm$  0.01 &   9.25 $\pm$  0.03 &   8.61 $\pm$  0.02 \\ 
     19 &  15.18 $\pm$  0.14 &  13.27 $\pm$  0.09 &  15.12 $\pm$  0.06 &  13.86 $\pm$  0.08 &  13.29 $\pm$  0.07 &  12.40 $\pm$  0.08 &  \dots &  \dots &  \dots \\ 
     20 &  15.33 $\pm$  0.15 &  13.31 $\pm$  0.09 & \dots &  \dots &  \dots &  \dots &  \dots &  \dots &  \dots \\ 
     21 &  \dots &  13.36 $\pm$  0.09 & \dots &  14.70 $\pm$  0.07 &  13.02 $\pm$  0.06 &  10.26 $\pm$  0.04 &   8.99 $\pm$  0.05 &   6.43 $\pm$  0.05 &  \dots \\ 
     22 &  \dots &  13.50 $\pm$  0.10 & \dots &  \dots &  12.94 $\pm$  0.06 &  11.20 $\pm$  0.04 &   9.76 $\pm$  0.02 &   8.63 $\pm$  0.02 &   7.57 $\pm$  0.02 \\ 
     23 &  \dots &  13.88 $\pm$  0.10 & \dots &  \dots &  13.57 $\pm$  0.06 &  11.89 $\pm$  0.08 &  11.18 $\pm$  0.10 &   9.79 $\pm$  0.15 &  \dots \\ 
     24 &  15.95 $\pm$  0.20 &  13.93 $\pm$  0.12 & \dots &  14.66 $\pm$  0.04 &  14.06 $\pm$  0.06 &  13.29 $\pm$  0.07 &  12.99 $\pm$  0.13 &  12.36 $\pm$  0.29 &  \dots \\ 
     25 &  16.63 $\pm$  0.31 &  14.03 $\pm$  0.13 & \dots &  \dots &  \dots &  \dots &  \dots &  \dots &  \dots \\ 
     26 &  15.95 $\pm$  0.20 &  14.06 $\pm$  0.12 &  15.80 $\pm$  0.06 &  14.76 $\pm$  0.05 &  14.24 $\pm$  0.07 &  13.91 $\pm$  0.07 &  13.64 $\pm$  0.11 &  \dots &  \dots \\ 
     27 &  16.90 $\pm$  0.33 &  14.30 $\pm$  0.14 & \dots &  15.07 $\pm$  0.07 &  14.09 $\pm$  0.06 &  12.62 $\pm$  0.05 &  12.64 $\pm$  0.12 &  \dots &  \dots \\ 
     28 &  \dots &  14.44 $\pm$  0.15 & \dots &  \dots & \dots &  12.82 $\pm$  0.18 &  12.45 $\pm$  0.20 &  \dots &  \dots \\ 
     29 &  \dots &  14.48 $\pm$  0.15 & \dots &  \dots &  \dots &  12.19 $\pm$  0.07 &  11.22 $\pm$  0.07 &   9.24 $\pm$  0.19 &   8.11 $\pm$  0.32 \\ 
     30 &  \dots &  14.49 $\pm$  0.16 & \dots &  \dots &  13.46 $\pm$  0.08 &   9.95 $\pm$  0.01 &   8.15 $\pm$  0.00 &   6.84 $\pm$  0.01 &  \dots \\ 
     31 &  \dots &  14.52 $\pm$  0.18 & \dots &  \dots &  \dots  &  12.13 $\pm$  0.03 &  11.41 $\pm$  0.02 &  \dots &   9.53 $\pm$  0.07 \\ 
     32 &  \dots &  14.60 $\pm$  0.19 & \dots &  \dots &  \dots &  12.69 $\pm$  0.14 &  11.51 $\pm$  0.09 &  11.03 $\pm$  0.28 &  \dots \\ 
     33 &  \dots &  14.65 $\pm$  0.20 & \dots &  \dots &  \dots &  \dots &  \dots &  \dots &   4.96 $\pm$  0.05 \\ 
     34 &  16.32 $\pm$  0.25 &  15.22 $\pm$  0.22 & \dots &  \dots &  \dots &  \dots &  \dots &  \dots &  \dots \\ 
     35 &  \dots &  15.59 $\pm$  0.24 & \dots &  \dots &  \dots &  \dots &  \dots &  \dots &  \dots \\ 
     36 &  \dots &  15.62 $\pm$  0.25 & \dots &  \dots &  \dots &  \dots &  \dots &  \dots &  \dots \\ 
     37 &  \dots &  15.70 $\pm$  0.25 & \dots &  \dots &  \dots &  13.40 $\pm$  0.10 &  12.46 $\pm$  0.07 &  12.46 $\pm$  0.37 &  \dots \\ 
     38 &  \dots &  15.85 $\pm$  0.26 & \dots &  \dots &  \dots &  \dots &  \dots &  \dots &  \dots \\ 
     39 &  \dots &  16.53 $\pm$  0.28 & \dots &  \dots &  \dots &  13.37 $\pm$  0.14 &  11.96 $\pm$  0.06 &  11.14 $\pm$  0.15 &  10.47 $\pm$  0.24 \\ 

%% file: tab2_16743.tex
VLA1 &6 10 50.50&-6 11 51.4	    &\dots&\dots& 1243		&VLA1& HII region$^\dagger$&\dots&+&\dots&\dots&\dots&\dots&+\\
     1 &  6 10 50.40  &  -6 11 19.3  & JK &  JHK  &  1234  &  $\dots$&  \textbf{early-B}  & V  &  + &  II   &  27700$\pm$5000  & 21.0$\pm$0.16 & 5.0$\pm$0.29 &+\\
      2 &  6 10 50.31  &  -6 11 57.6  & JK &  \dots  &  1234  &  VLA 4 & Herbig Be$^{\dagger\dagger}$  & $\dots$ &  + &  0 / I   &  \dots  & \dots & \dots&+\\

      3 &  6 10 49.81  &  -6 11 45.0  & JK &  JHK  &  1234  & \dots &  G0 - G3  & V  &  + &  0 / I   &  5950$\pm$500  & 10.9$\pm$0.17 & 4.2$\pm$0.25&+\\
      4 &  6 10 49.45  &  -6 11 42.0  & K &  HK  &  1234  & VLA 2 &  \dots  &  $\dots$  &  + &  0 / I   &  \dots  & \dots & \dots&+\\
      5 &  6 10 50.50  &  -6 12 04.4  & JK &  JK  &  1  &  $\dots$ &  K1 - K2  &  IV  &  \dots &  $\dots$   &  5000$\pm$350  & 4.7$\pm$0.52 & 3.4$\pm$0.21 &\dots \\
      6 &  6 10 50.68  &  -6 11 55.1  & JK &  JK  &  1  &  $\dots$ &  K2  &  IV  &  + &  $\dots$   &  4900$\pm$350  & 12.9$\pm$0.25 & 3.7$\pm$0.27&+\\
      7 &  6 10 49.21  &  -6 11 30.2  & JK &  JHK  &  12  &  VLA 8 & K1  &  IV  &  no &  $\dots$   &  5100$\pm$350  & 8.2$\pm$0.45 & 2.3$\pm$0.14 &+\\
      8 &  6 10 52.52  &  -6 11 35.0  & JK &  K  &  1  &  $\dots$ & K0 - K2  &  IV  &  + &  $\dots$   &  5100$\pm$350  & 12.5$\pm$0.80 & 2.9$\pm$0.18 &+\\
      9 &  6 10 48.63  &  -6 12 00.4  & JK &  JHK  &  1234  &  $\dots$ &  G8 - K0  &  IV  &  + &  II   &  5400$\pm$350  & 13.6$\pm$0.74 & 3.0$\pm$0.18 &+\\
     10 &  6 10 49.65  &  -6 11 22.1  & JK &  K  &  1  &  $\dots$ & K1  &  IV  &  \dots &  $\dots$   &  5100$\pm$350  & 7.5$\pm$0.45 & 2.1$\pm$0.12 &\dots \\
     11 &  6 10 50.82  &  -6 11 39.6  & K &  HK  &  123  &  $\dots$ &  \dots  &  \dots  &  + &  $\dots$   &  \dots  & \dots & \dots &+\\
     12 &  6 10 50.61  &  -6 11 35.0 & K &  HK  &  123 &  \dots& $\dots$  &  $\dots$  &  + &  $\dots$   &  \dots  & \dots & \dots &+\\
     13 &  6 10 48.18  &  -6 11 25.7  & JK &  JHK  &  1234 &  $\dots$&  K1  &  V  &  + &  II   &  5100$\pm$350  & 8.3$\pm$0.20 & 1.9$\pm$0.11 &+\\
     14 &  6 10 50.39  &  -6 11 55.3  & JK &  \dots  &  $\dots$  &  $\dots$ &M4  &  V  &  \dots &  $\dots$   &  3350$\pm$350  & 4.9$\pm$0.42 & 2.4$\pm$0.15 &no\\
     15 &  6 10 47.74  &  -6 12 00.9  & JK &  HK  &  1234  &  \dots & K5  &  IV  &  + &  II   &  4350$\pm$250  & 13.7$\pm$0.61 & 2.3$\pm$0.13 &+\\
     16 &  6 10 50.28  &  -6 12 05.6  & JK & \dots  &  $\dots$  &  $\dots$ &K5  &  IV  &  \dots &  $\dots$   &  4350$\pm$250  & 7.8$\pm$0.61 & 2.3$\pm$0.13 &\dots \\
     17 &  6 10 48.82  &  -6 11 43.8  & JK &  JHK  &  1234  &  $\dots$ & K1  &  IV  &  + &  II   &  5100$\pm$250  & 7.8$\pm$0.45 & 1.4$\pm$0.08 &+\\
     18 &  6 10 48.78  &  -6 11 32.6  & K &  K  &  1234  &  $\dots$ &  $\dots$  &  $\dots$  &  + &  0 / I   &  \dots  & \dots & \dots &+\\
     19 &  6 10 49.64  &  -6 11 15.3 & JK &  JHK  &  1  &  $\dots$ & K1  &  IV  &  \dots &  $\dots$   &  5100$\pm$350  & 6.2$\pm$0.45 & 1.2$\pm$0.07 &+\\
     20 &  6 10 50.12  &  -6 11 59.0  & JK &  \dots  &  $\dots$  &  $\dots$ &  $\dots$  &  $\dots$  &  \dots &  $\dots$   &  \dots  & \dots & \dots&+\\
     21 &  6 10 50.76  &  -6 11 44.9  & K & HK  &  123  &  $\dots$ &K3 - K4  & IV  &  + &  $\dots$   &  4650$\pm$350  & 23.9$\pm$0.24 & 2.1$\pm$0.13 &+\\
     22 &  6 10 49.44  &  -6 11 20.6  & K &  K  &  1234  &  $\dots$ &$\dots$  &  $\dots$  &  + &  0 / I   &  \dots  & \dots & \dots &+\\
     23 &  6 10 51.49  &  -6 11 35.0  & K &  HK  &  123 &  $\dots$ &$\dots$  &  $\dots$  &  + &  $\dots$   &  \dots  & \dots & \dots &+\\
     24 &  6 10 48.02  &  -6 11 42.8  & JK &  HK  &  123  &  $\dots$ &  \dots  &  \dots  &  + &  $\dots$   &  \dots  & \dots & \dots &+\\
     25 &  6 10 50.16  &  -6 11 55.0  & JK &  \dots  &  $\dots$  &  $\dots$ & $\dots$  &  $\dots$  &  \dots &  $\dots$   &  \dots  & \dots & \dots &+\\
     26 &  6 10 48.15  &  -6 11 06.6  & JK &  JHK  &  12  &  $\dots$ & $\dots$  &  $\dots$  &  \dots &  $\dots$   &  \dots  & \dots & \dots &\dots \\
     27 &  6 10 52.01  &  -6 12 12.3  & JK &  HK  &  12  &  $\dots$ & $\dots$  &  $\dots$  &  \dots &  $\dots$   &  \dots  & \dots & \dots &\dots \\
     28 &  6 10 51.10  &  -6 11 25.3  & K &  HK  &  12  &  $\dots$ & $\dots$  & $\dots$  &  \dots &  $\dots$   &  \dots  & \dots & \dots &\dots \\
     29 &  6 10 51.64  &  -6 12 07.8  & K &  K  &  1234  &  $\dots$ & \dots  &  \dots  &  + &  0 / I   &  \dots  & \dots & \dots &+\\
     30 &  6 10 52.35  &  -6 11 32.1  & K &  K  &  123  &  VLA 7 & \dots  &  \dots  &  + &  $\dots$   &  \dots  & \dots & \dots &+\\
     31 &  6 10 48.43  &  -6 12 03.9  & K &  HK  &  124  &  $\dots$ & \dots  &  \dots  &  \dots &  $\dots$   &  \dots  & \dots & \dots &+\\
     32 &  6 10 51.55  &  -6 11 26.4  & K &  K  &  123  &  $\dots$ & \dots  &  \dots  &  + &  $\dots$   &  \dots  & \dots & \dots & \dots \\
     33 &  6 10 50.75  &  -6 11 42.0  & K &  \dots  &  4  & VLA 6 & \dots  &  \dots  &  \dots &  $\dots$   &  \dots  & \dots & \dots &+ \\
     34 &  6 10 52.23  &  -6 11 25.2  & JK &  \dots  &  $\dots$  &  $\dots$ & \dots  &  \dots  &  \dots &  $\dots$   &  \dots  & \dots & \dots &\dots \\
     35 &  6 10 50.54  &  -6 11 19.0  & K &  \dots  &  $\dots$  &  $\dots$ & \dots  &  \dots  &  \dots &  $\dots$   & \dots  & \dots & \dots &\dots \\
     36 &  6 10 50.98  &  -6 11 38.3  & K &  \dots  &  $\dots$  &  $\dots$ & \dots  &  \dots  &  \dots &  $\dots$   &  \dots  & \dots & \dots &\dots \\
     37 &  6 10 47.90  &  -6 11 47.9  & K &  \dots  &  123   &\dots& \dots  &  \dots  &  + &  $\dots$   &  \dots  & \dots & \dots & + \\
     38 &  6 10 49.31  &  -6 11 46.3  & K &  \dots  &  $\dots$  &\dots& \dots  &  \dots  &  \dots &  $\dots$   &  \dots  & \dots & \dots &\dots \\
     39 &  6 10 51.80  &  -6 11 16.3  & K &  \dots  &  1234  &  $\dots$ & \dots  &  \dots  &  + &  0 / I   &  \dots  & \dots & \dots &+\\